\documentclass[useAMS,usenatbib]{mn2e}
\usepackage{amsmath,amsfonts,bbm, graphicx,color}
\usepackage{aas_macros}
\usepackage{url}
\usepackage{multirow}
\def\be{\begin{equation}}
\def\ee{\end{equation}}
\def\ba{\begin{eqnarray}}
\def\ea{\end{eqnarray}}

\def\k{\mathbf{k}}
\def\<{\langle}
\def\>{\rangle}

\title[Nonlinear stochastic growth rates and redshift space distortions]{ Nonlinear stochastic growth rates and redshift space distortions} %
\author[E. Jennings \& D. Jennings]{ Elise Jennings$^{1,2}$\thanks{E-mail: elise@fnal.gov} \& David Jennings$^{3}$\\ 
$^{1}$Center for Particle Astrophysics, Fermi National Accelerator Laboratory MS209, P.O. Box 500, Kirk Rd. \& Pine St., Batavia, IL 60510-0500\\
$^{2}$Kavli Institute for Cosmological Physics, Enrico Fermi Institute, University of Chicago, Chicago, IL 60637\\
$^{3}$Controlled Quantum Dynamics Theory, Department of Physics, Imperial College London, London SW7 2AZ,United Kingdom
}

\begin{document}

\date{}


\maketitle


\begin{abstract}
The linear growth rate is commonly defined through a simple deterministic relation between the velocity divergence and the matter overdensity in the linear regime. 
We introduce a formalism that extends this to a nonlinear, stochastic relation between $\theta = \nabla \cdot v({\bf x},t)/aH$ and $\delta$. This provides a new phenomenological approach that examines the conditional mean $\langle \theta|\delta\rangle$, together with the fluctuations of $\theta$ around this mean.
We measure these stochastic components using N-body simulations and find they are non-negative and increase with decreasing scale from $\sim$10\%  at $k<0.2 h $Mpc$^{-1}$ to 25\% at $k\sim0.45h$Mpc$^{-1}$  at $z = 0$.  
Both the stochastic relation and nonlinearity are more pronounced for halos, $M \le 5 \times 10^{12}M_\odot h^{-1}$, compared to the dark matter at $z=0$ and $1$. 
Nonlinear growth effects manifest themselves as a rotation of the mean $\langle \theta|\delta\rangle$ away from the linear theory prediction
$-f_{\tiny \rm LT}\delta$, where $f_{\tiny \rm LT}$ is the linear growth rate. This rotation increases with wavenumber, $k$, and we show that it can be well-described by second order Lagrangian perturbation theory (2LPT) for $k < 0.1 h$Mpc$^{-1}$.
The stochasticity in the $\theta$ -- $\delta$ relation is not so simply described by 2LPT, and we discuss its impact on measurements of  $f_{\tiny \rm LT}$ from two point statistics in redshift space. Given that the relationship between $\delta$ and $\theta$ is stochastic and nonlinear, this will have implications for the interpretation and precision of  $f_{\tiny \rm LT}$ extracted using models which assume a linear, deterministic expression. 
\end{abstract}


\section{Introduction}

The clustering of galaxies on Mpc scales in the Universe is a fundamental cosmological observable which
allows us to constrain key parameters of the  $\Lambda$CDM model and to look for deviations from this standard model.
Understanding the relationship between peculiar velocity flows and the large scale mass distribution is crucial to interpreting the clustering signal measured in redshift space, where these velocities distort
 the clustering amplitude along the line of sight \citep[see e.g.][]{2001Natur.410..169P,2008Natur.451..541G,
2011MNRAS.415.2876B,2012MNRAS.426.2719R,2014MNRAS.443.1065B}.
In this paper we investigate the assumptions of a linear and deterministic relation
between the peculiar velocity and overdensity fields at a range of scales and redshifts. We present a general formalism
where deviations from linearity and determinism can be viewed separately in the two point clustering statistics of the
velocity divergence auto and cross power spectra. This approach
represents a new phenomenological tool based on a stochastic description of nonlinear effects.

One of the key aims of future galaxy redshift surveys
\citep{Cimatti:2009is,2013arXiv1305.5422S,2015AAS...22533605E}
is to measure this linear perturbation theory relation between the density and velocity fields,
referred to as the linear growth rate, to less than 1\% precision using the redshift space clustering statistics of different galaxy tracers.
This level of accuracy has motivated a lot of work in developing a precise model for the two point clustering statistics either as the correlation
function in configuration space \citep[e.g.][]{2011MNRAS.417.1913R,2015MNRAS.446...75B} 
or the power spectrum in Fourier space \citep[e.g][]{1994MNRAS.267.1020P,Scoccimarro:2004tg,2011MNRAS.410.2081J,
2011JCAP...11..039S, 2013PhRvD..87h3509T}.
Note that many of these studies are based on a mix of assumptions of either a
linear and/or deterministic density velocity relation.

Current models for the two point clustering statistics in redshift space that include
perturbation theory expansions have been shown to be an improvement over linear theory
 in modelling these redshift space clustering statistics.
Although all are limited to very large scales $k<0.15h$Mpc$^{-1}$ at low redshifts \citep[see e.g][]{Scoccimarro:2004tg, 2011MNRAS.410.2081J, 2012ApJ...748...78K} and moreover may
 only apply to highly biased tracers \citep{2011MNRAS.417.1913R}; none of the models can recover the linear growth rate to a
percent level accuracy on the scales which will be probed by future galaxy surveys.
If we are to limit our analysis of redshift space distortions to large scales, where quasi-linear theory models apply,
then it is worthwhile investigating both
where the  assumptions of a linear and deterministic relation between the density and velocity fields breaks down and
how well perturbation theory expansions can recover these components.

This formalism involving the decomposition of the two point statistics into nonlinear and stochastic
components is both well defined and consistent with a full perturbation theory expansion of all the nonlinear effects. The approach provides an alternative, more phenomenological description of such nonlinear effects. In considering either the galaxy - dark matter overdensity relation or the velocity- overdensity relation, there
 is a general notion of stochasticity which is often not well defined and or vaguely explained as due to a nonlinear coupling of modes. In this paper, our use of the term stochasticity refers to the break-down of a deterministic relation that exists in the linear regime between the growing overdensity field and the velocity divergence. We also discuss the connection between such a notion of stochasticity and mode coupling in standard perturbation theory.

It is well known that the halo or galaxy overdensity field does not trace the dark matter field faithfully and that the relation between the two is generally described by a linear bias term which is scale independent and is different for different galaxy tracers \citep[see e.g][]{1999ApJ...520...24D,1999ApJ...520..437K}. Recently there is renewed interest in considering the stochasticity in this relation on large scales  \citep{2004MNRAS.355..129S,2009MNRAS.396.1610B,2013PhRvD..87l3523S} where previously we would have assumed a linear, deterministic relationship to hold. Also, as noted in \citet{2004MNRAS.355..129S}, dominant perturbative corrections come from mode coupling at wavelengths close to the wavelength of the mode itself. Long wavelength modes sampled from a finite volume can have significant fluctuations which would give rise to significant fluctuations in second order corrections.

There have been many studies that have compared the  two point statistics of the matter and velocity divergence fields and found them
to be nonlinear on large scales ($k\sim 0.1 h/$Mpc) which are traditionally considered the linear regime
 \citep[][]{Scoccimarro:2004tg, 2009MNRAS.393..297P, 2011MNRAS.410.2081J, 2012MNRAS.427L..25J,2012MNRAS.427.2537C, 2012MNRAS.425.2128J}.  \citet{2012MNRAS.427L..25J} measured this nonlinearity as the deviation of the velocity divergence power spectra
$P_{\theta \theta} :=\langle \theta({\bf k}) \theta^*({\bf k'}) \rangle$
 and
 $P_{\theta \delta} :=\langle \theta({\bf k}) \delta^*({\bf k'}) \rangle$ from linear perturbation theory predictions 
 and found it to be at the level of 20\% and 10\% respectively at $k \sim 0.1 h$Mpc$^{-1}$.
Note that these nonlinear features are at the level of the ensemble averaged two point statistics.
In contrast, in this work we will examine the velocity divergence -- overdensity relation,  $\theta - \delta$, in Fourier space at each wavenumber where we can separate the notion of nonlinear and stochastic effects.

 \citet{1999MNRAS.309..543B} investigated the statistical relation between the density and velocity fields in the mildly nonlinear regime, focusing on the conditional
probability distribution $P(\theta|\delta)$ of the smoothed fields in configuration space. This study of the stochastic relationship between the two fields used simulations of a small volume, (200Mpc$/h$)$^3$, and low resolution,128$^3$ particles, by today's standards. Given the high resolution and large volume simulations we have available today and our knowledge of how sensitive velocity statistics are to resolution effects \citep{2009PhRvD..80d3504P, 2011MNRAS.410.2081J, 2015MNRAS.446..793J,2014arXiv1410.1256Z,2014PhRvD..90j3529B} it is important to revisit this study.
In this paper we explore a formalism that defines both a nonlinear and a stochastic relation between the velocity divergence and the conditional mean value of this function at a given overdensity. We also investigate the variance of the velocity divergence around this relation as a function of scale, which
defines a stochastic description of nonlinear effects.

The paper is laid out as follows:
In Section \ref{sec:nbody} we describe the N-body simulations and tessellation techniques used to measure both the density and
velocity divergence fields of dark matter and halos in this paper.
In Section \ref{sec:lin} we present the linear perturbation theory relation between the density and velocity fields.
In Sections \ref{sec:nonlinear}  we outline the main formalism in this paper which defines the nonlinearity and the stochastic relation
between the velocity divergence and overdensity fields and  give expressions for the two point statistics.
In Section \ref{sec:results} we present our results.
The measurement of the conditional mean relation and scatter about this mean are presented in Sections
\ref{sec:nl_f} and \ref{sec:stoc} for dark matter and in Section \ref{sec:halos} for halos. 
In Section \ref{sec:pt} we relate the two point functions in this paper to both one loop standard
perturbation and second order Lagangian perturbation theory predictions.
In Section \ref{sec:rsd} we discuss the impact of a nonlinear and stochastic relation between the velocity and density fields on
models for the power spectrum in redshift space.
In Section \ref{sec:conc} we summarize our results.

\section{Density and velocity two point statistics from N-body simulations}
\label{sec:nbody}
In section \ref{sec:sims} we present the details of the dark matter N-body simulations and the MultiDark halo catalogue used in this work.
In  Section \ref{measuring_v} we
outline the methods used to measure both the velocity divergence power spectrum 
and the matter power spectrum as a function of scale.

\subsection{N-body simulations}
\label{sec:sims}

We use the N-body simulations carried out by \citet{2012JCAP...01..051L, 2013MNRAS.428..743L}.
These simulations were performed using
a  modified version of the mesh-based $N$-body code {\tt RAMSES} \citep{2002A&A...385..337T}.
Assuming  a $\Lambda$CDM cosmology,
the following cosmological parameters were used in the simulations:
$\Omega_{\rm m} = 0.24$,
 $\Omega_{\rmn{DE}}=0.76$,
$h = 0.73$ and a spectral tilt of $n_{\mbox{s}} =0.961$ \citep[in agreement with e.g.][]{Sanchez:2009jq}.
The  linear theory rms fluctuation
in spheres of radius 8 $h^{-1}$ Mpc is set to be  $\sigma_8 = 0.769$.
The simulations use $N=1024^3$ dark matter  particles to represent the  matter distribution in a  computational box of
comoving length $1500 h^{-1}$Mpc.
The initial conditions were generated at $z=49$ using the  MPgrafic\footnote{http://www2.iap.fr/users/pichon/mpgrafic.html} code. The errors on the power spectra in this work are calculated from the variance in the two point statistics from six simulations of the same cosmology initialized with different 
realizations of the dark matter density field.

We use the publicly
available halo catalogues from the MultiDark 
simulation \citep{ 2011arXiv1109.0003R, 2012MNRAS.423.3018P} which has a computational box size of L =
1000$h^{-1}$Mpc  on a side.
These halos have been identified using the Bound-Density-Maxima algorithm \citep{1997astro.ph.12217K}. 
The halo sample we use in this work consists of all haloes with  $M \le 5 \times 10^{12}h^{-1}M_\odot$ at $z=0$ and $z=1$.
The error on the halo power spectrum in a spherical shell of width $\delta k$ is estimated 
using the following formula derived by  \citet{1994ApJ...426...23F}:
\begin{equation}
\frac{\sigma}{P} = \sqrt{\frac{(2\pi)^2}{V k^2\delta k}}\left( 1+ \frac{1}{P\bar{n}}\right) ,
\label{eq:pkerror}
\end{equation}
where $\bar{n}$ is the number density and $V$ is the volume.
We measure the linear bias, $b$, for this sample of halos by fitting to the ratio $b = \sqrt{\langle \delta_H \delta_H^*\rangle/\langle \delta_{\tiny \rm LT} \delta_{\tiny \rm LT}^*\rangle}$
 on large scales $k<0.1h$Mpc$^{-1}$, where $\delta_H$ is the nonlinear halo overdensity in Fourier space. Here $\langle \delta_{\rm \tiny LT} \delta_{\tiny \rm LT}^*\rangle$ is 
 the $z=0$ linear theory power spectrum 
generated using CAMB with the same cosmological parameters used in the MultiDark simulations.

\subsection{Measuring the density and velocity fields \label{measuring_v}}

The nonlinear matter and halo power spectra are measured from the simulations
by assigning the particles to a mesh using the cloud in cell (CIC) assignment scheme  \citep{1988csup.book.....H} onto a 512$^3$ grid and
performing a fast Fourier transform (FFT) of the density field.
To compensate for the mass assignment scheme we perform an approximate de-convolution following  \citet{1991ApJ...375...25B}.

Measuring the velocity divergence field accurately from numerical simulations on small scales
can be difficult if a mass weighted approach is used as in \citet[][]{Scoccimarro:2004tg,2009PhRvD..80d3504P, 2011MNRAS.410.2081J}.
Some volume weighted measures of the velocity field have also been developed \citep[see e.g.][]{1996MNRAS.279..693B, 2007MNRAS.375..348C} including the
Delaunay tessellation field estimator (DTFE) method \citep{2007PhDT.......433S, 2011ascl.soft05003C}.

In the mass weighted approach, simply interpolating the velocities
to a grid, as suggested by \citet{Scoccimarro:2004tg}, gives the momentum field which is then Fourier
transformed and divided by the Fourier transform of the density field, which results in a mass weighted velocity field on the grid.
One of the main problems with this approach is that the velocity field is artificially set to zero in regions where
there are no particles, as the density is zero in these empty cells.
\citet{2009PhRvD..80d3504P} also found that this method does not accurately recover the
input velocity divergence power spectrum on scales $k>0.2 h$Mpc$^{-1}$ interpolating the velocities of 640$^3$ particles to a $200^3$ grid.
Using simulations of 1024$^3$ particles in a 1.5$h^{-1}$Gpc box, \citet{2011MNRAS.410.2081J}  found that the maximum
grid size that could be used was 350$^3$ without reaching the limit of empty cells.

In this paper the velocity divergence fields are measured from the N-body simulations using the
DTFE method  \citep{2007PhDT.......433S, 2011ascl.soft05003C}.
This code  constructs the Delaunay tessellation from a discrete set of points and interpolates the field values onto a user defined grid.
For the $L_{\tiny \mbox{box}} = 1500h^{-1} $ Mpc simulation
we  generate all two point statistics on a $512^3$ grid. We have verified that our results do not change when we increase
the grid size to 1024$^3$, demonstrating 
that our two point clustering statistics have converged on the relevant scales in this paper.
The velocity divergence field is interpolated onto the
grid by  randomly sampling the field values at a given number of sample points within the Delaunay cells and
than taking the average of those values.
The resolution of the mesh used in this study means that
mass assignment effects are negligible on the scales of interest here.
Throughout this paper the velocity divergence is normalized to a dimensionless quantity $\theta  = -\nabla \cdot v /(a H )$,
where $v$ is the peculiar velocity, $H$ is the Hubble parameter and $a$ is the scale factor.

It has recently been shown that there exists a non-negligible velocity bias on large scales between the halo and dark matter velocity fields. This is a statistical manifestation of sampling effect which increases with decreasing number density \citep[see e.g.][]{2014PhRvD..90j3529B, 2014arXiv1410.1256Z, 2015MNRAS.446..793J}. We use haloes of mass $M \le 5 \times 10^{12}h^{-1}M_\odot$ from the MultiDark simulation which have a number density of $\bar{n} = 1.23\times 10^{-2} ($Mpc$/h)^{-3}$ at $z=0$ so that the velocity bias is negligible on the relevant scales discussed in this paper. 
Note that certain methods of measuring either the
velocity or velocity divergence field, e.g. the nearest grid point method, can induce extra sampling effects in addition to the statistical bias mentioned above \citep[see e.g.][]{2014arXiv1405.7125Z}, the DTFE method does not suffer from the same sampling effects \citep{2007PhDT.......433S} and will not impact our analysis
which is restricted to scales $k<0.45 h$Mpc$^{-1}$.

\section{ The density -- velocity field relation}
\subsection{ Linear theory}
\label{sec:lin}

\begin{figure*}
\begin{center}
\includegraphics[height=4.5in,width=4.5in]{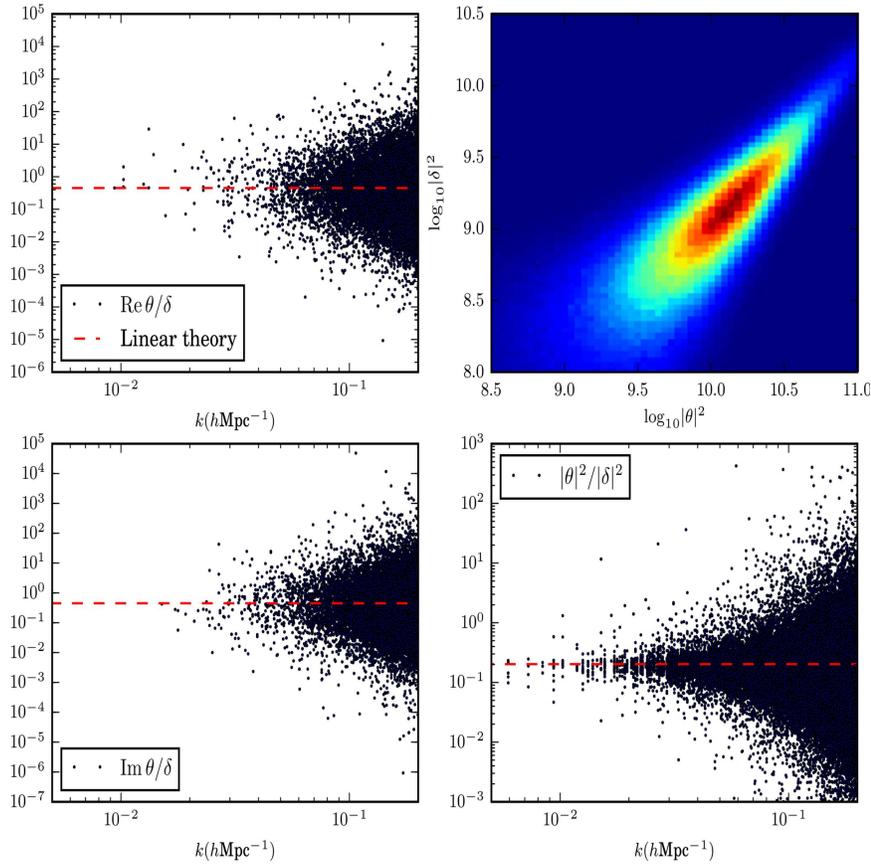}
\caption{The upper and lower left panels show the ratio of Re$(\theta({\bf k})/\delta({\bf k}))$  (Im($\theta({\bf k})/\delta({\bf k})$)) as a function of wavenumber, $k$, measured from the simulations at $z=0$.
The joint pdf ${\mathcal P} ({\rm log}|\theta|^2, {\rm log}|\delta|^2)$ is plotted in the top right panel. The ratio of the magnitudes $|\theta|^2/|\delta|^2$ for each mode  is plotted in the lower right panel as a function of scale. In all panels the linear theory prediction for the $\theta - \delta$ relation is shown as a red dashed line. \label{fig:1}}
\end{center}
\end{figure*}

At large scales the Universe is homogeneous and the fluctuation fields $\delta({\bf x},t) = \rho({\bf x},t)/\bar{\rho} -1, {\bf v({\bf x},t)}, \Phi({\bf x},t)$ are small compared to the smooth background contributions. An Eulerian approach to density fluctuations relies on a truncation of the full Vlasov equation and the imposition of an equation of state. Under the assumption of zero shear, the linear regime is then described by the continuity and Euler equations,
\begin{align}
\frac{\partial \delta({\bf x},\tau)}{\partial \tau} +  {\bf \nabla} \cdot {\bf v}({\bf x},\tau) &= 0 \\
\frac{\partial {\bf v}({\bf x},\tau)}{\partial \tau} + aH {\bf v}({\bf x},\tau) &= -{\bf \nabla} \Phi({\bf x},\tau)\, ,
\end{align}
where ${\rm d}t =a {\rm d}\tau$.
The {{linear theory growth rate}}, $f_{\tiny \rm LT}$ is defined as the logarithmic derivative of the overdensity field, and is dependent on the cosmological parameters,
\begin{equation}
f_{\tiny \rm LT}(\Omega_m, \Omega_\Lambda) := \frac{{\rm d ln}\delta}{{\rm d ln}a} \,.
\end{equation}
The growing mode solution for $\delta({\bf x},t )$ admits a product form in which it separates 
as $\delta ({\bf x},t) = D(t)\delta({\bf x},0)$, where $D$ is the linear growth factor. For this product form the
linear growth rate becomes the logarithmic derivative of the growth factor, $f_{\tiny \rm LT}(t)  = {\rm d ln}D(t)/{\rm d ln}a$.

Together with the linear continuity equation, we find that the velocity divergence and overdensity fields are simply related as
\begin{equation}
\theta({\bf x},t) := \frac{\nabla \cdot v({\bf x},t)}{aH} = -f_{\rm \tiny LT}(\Omega_m, \Omega_\Lambda) \delta({\bf x},t)
\label{eqn.linearcontinuity}
\end{equation}
where we define $\theta$ as the velocity divergence in units of $(aH)$ and $\bf{v}({\bf x},t)$ is the comoving peculiar velocity. Since we are within the linear regime, this relation carries over trivially to Fourier space, where
$\theta({\bf k},t) = -f_{\rm \tiny LT} \delta({\bf k},t)$. Put another way, the linear regime is special in that it admits the introduction of a linear growth rate $f_{LT} (\Omega_m, \Omega_\Lambda)$ that is independent of the scale at which we measure the perturbations.

However, we do not expect this relation to hold once the density fluctuations in the fields become large, and non-linear growth starts to generate mode-coupling. In what follows we shall analyse to what extent it is possible to sensibly extend the central relation Eq. (\ref{eqn.linearcontinuity}) beyond the linear regime, and to provide meaningful insights into bulk characteristics that arise from non-linearities. We find that the relation is modified in essentially two ways: firstly one finds a growth factor that is scale dependent due to non-linearities, and secondly we find that the deterministic one-to-one relation between $\theta$ and $\delta$ is weakened to a stochastic relation. In section \ref{sec:pt} we describe how these results are understood from the perspective of perturbation theory, and in section \ref{sec:rsd} discuss implications for redshift space distortions.

\subsection{ A nonlinear stochastic relation between the density and velocity fields }
\label{sec:nonlinear}

Random fields in cosmology are used to represent a single realization of the dark matter distribution within a given cosmology. As these fields evolve under gravity, nonlinearities 
give rise to a growth in structure which induce correlations between different scales. The full nonlinear equations of motion in Fourier space are given by
\begin{align}
&\frac{1}{aH}\partial_\tau\delta({\bf k},\tau) +  \theta({\bf k},\tau) = - \int{\rm d}^3{\bf k}_1  A({\bf k}_1, {\bf k} -{\bf k}_1) \theta({\bf k}_1)\delta({\bf k} - {\bf k}_1)  \label{eq:nl1}\\
&\partial_\tau \theta({\bf k},\tau) + aH \theta({\bf k},\tau) \nonumber \\
&+ \frac{3}{2}\Omega_{\rm m} aH\delta({\bf k},\tau)= - \int{\rm d}^3{\bf k}_1  B({\bf k}_1, {\bf k} - {\bf k}_1)\theta({\bf k}_1)\theta({\bf k} - {\bf k}_1)\, ,
\label{eq:nl2}
\end{align}
where we have the mode-coupling functions 
\begin{align}
A(\k_1, \k_2) &= \frac{(\k_1 + \k_2) \cdot \k_1}{k_1^2} \nonumber \\
B(\k_1, \k_2) & = \frac{|\k_1 + \k_2|^2(\k_1\cdot \k_2)}{2k_1^2k_2^2}.
\end{align}
The terms on the right hand side of both Eq. (\ref{eq:nl1}) and (\ref{eq:nl2}) encode the nonlinear evolution of the fields \citep[see e.g.][ for a review]{2002PhR...367....1B}.
Computing the perturbative components for these nonlinear contributions is an on-going challenge, and the complexity increases rapidly with higher order terms \citep[see e.g.][]{2006PhRvD..73f3519C,2008PhRvD..77b3533C}. In what follows we construct a phenomenological approach to describe the breakdown of the linear theory relationship in simple terms. 

As mentioned, the departure point from linear theory that provides our focus is the relationship between the overdensity field and the velocity divergence. To illustrate this breakdown  the upper (lower) left panel of Figure \ref{fig:1} shows the scatter in the ratio of $\rm{Re}[\theta(\k)/\delta(\k)]$ and $\rm{Im}[\theta(\k)/\delta(\k)]$ as a function of wavenumber, $k$, measured from the simulations at redshift $z=0$. 

Significant scatter exists in the Fourier modes about the linear theory relation (red dashed line), and which increases as a function of scale. In the lower right panel of this figure we also plot the ratio of the magnitudes $|\theta(\k)|/|\delta(\k)|$ as a function of scale, which demonstrates that this scatter is not due to an arbitrary phase differences between the modes, and which could have cancelled when computing the two point statistics of the fields. The scatter in the $\theta - \delta$ relation is also shown in the upper right panel of Fig \ref{fig:1}, where we plot the PDF of the logarithm of $|\theta^2|$ and $|\delta^2|$. 

Firstly, we introduce a conditioned velocity divergence quantity $\<\theta | \delta\>$ that is derived from the conditional distribution $P(\theta | \delta)$. More explicitly, we define $\<\theta (\k, t) | \delta\> := \int \mathcal{D}\theta \, P(\theta |\delta) \theta(\k, t)$ for the conditional expectation value of $\theta$. The resultant term has a dependence on the particular overdensity that is realised. In the linear regime a direct relation exists between $\theta$ and $\delta$, and corresponds to a delta function distribution in $P(\theta, \delta)$ for which $P(\theta |\delta)$ is perfectly sharp, or ``deterministic'', and encodes the relation $\theta = -f_{LT} \delta$. However we can extend this to a more general scenario that drops this sharp relation in favor of a stochastic one. We define a growth rate $f_\delta(\Omega_m, \Omega_\Lambda, \k)$, in momentum space, as 
\begin{equation}
f_\delta (\Omega_m, \Omega_\Lambda, \k) := -\frac{1}{\delta( \k, t) } \< \theta (\k ,t ) | \delta \>_{\theta|\delta}.
\end{equation}
 Here the generalized growth rate now has an explicit dependence on the overdensity field that is being conditioned on, in addition to a potential scale dependence. Importantly, in the linear regime this function coincides with the linear growth rate $f_{LT}$, but more generally becomes a stochastic quantity for which moments can be computed.
 
To estimate the non-linear distortions to the effective growth rate, it is instructive to compute the following moments
\begin{align}
\hat{f} &:= \frac{\langle \langle \theta |\delta\rangle \delta\rangle}{\langle \delta^2\rangle} = \frac{-\langle  f_\delta\delta^2 \rangle_{\delta}}{\langle \delta^2\rangle}  \label{eq:fvariables1} \\ 
\tilde{f}^2 &:= \frac{\langle \langle \theta |\delta\rangle \langle \theta|\delta\rangle\rangle}{\langle \delta^2\rangle} = \frac{\langle f^2_\delta \delta^2 \rangle_{\delta}}{\langle \delta^2\rangle},  \label{eq:fvariables2}
\end{align}
where by definition, $\langle \delta \rangle = 0$. Here $\langle \cdot \rangle_\delta$ denotes an ensemble average with respect to the probability distribution function $P(\delta)$, however from now on we will omit the subscript from any ensemble average notation, for simplicity.
In the linear regime we automatically have that $|\hat{f}| = |\tilde{f}| = f_{\rm LT}(\Omega_m, \Omega_\Lambda)$, as expected.

In addition to these non-linear distortions to $f$, we recall that the essential connection between $\theta$ and $\delta$ gradually becomes diluted to a stochastic one. This can be quantified through the fluctuations of $\theta(\k)$ about the conditional expectation. In particular, we consider the following random field
\begin{equation}
\alpha(\k,t) := \theta({\bf k},t) - \langle \theta({\bf k},t) | \delta \rangle
\label{eq:alpha}
\end{equation}
whose variance provides a suitable measure, and is given by
\begin{equation}
\sigma_\alpha^2 = \frac{\langle \alpha^2\rangle}{\langle \delta^2\rangle}  = \frac{\langle \theta^2 \rangle - \langle \langle \theta |\delta \rangle^2\rangle}{\langle \delta^2\rangle}.
\end{equation}
Again if the linear continuity equation, Eq. (\ref{eqn.linearcontinuity}), holds then $\sigma_\alpha = 0$ and stochastic relation vanishes. Section \ref{sec:results} contains a closer examination of the nonlinear moments in  Eqs. (\ref{eq:fvariables1}) and (\ref{eq:fvariables2}), and the magnitude of $\alpha$ measured from the N-body simulations as a function of both scale and redshift.

More generally it is seen that the quantity $\hat{f}$ is related to the expected velocity divergence at a particular scale through the relation 
\begin{align}
\<\theta\> =-\int {\rm d}^3k_1 &A(\k_1, \k-\k_1)  \times \\   \nonumber
&\left [\hat{f}\< \delta(\k_1) \delta(\k-\k_1) \> +  \< \alpha(\k_1) \delta(\k-\k_1)\>
 \right ]
\end{align}
which follows from the full non-linear continuity equation. A parallel relation for $\tilde{f}$ can be obtained, and from the Euler equation we find
\begin{align}
\< \dot{\theta}\>  &= - \int \rm{d}^3k_1 \\ \nonumber
& \left [ B(\k_1, \k - \k_1) (\tilde{f}^2\langle \delta({\bf k_1}) \delta({\bf k} - {\bf k_1}) \rangle +  \langle \alpha({\bf k_1}) \alpha({\bf k} - {\bf k_1}) \rangle) \right .  \\ \nonumber
&\left . - A(\k_1, \k-\k_1) (\hat{f}\langle \delta({\bf k_1}) \delta({\bf k} - {\bf k_1}) \rangle +  \langle \alpha({\bf k_1}) \delta({\bf k} - {\bf k_1}) \rangle )\right ] \, ,
\end{align}
where $\alpha$ quantifies the deviation from a deterministic relation between $\theta$ and $\delta$ and we have used the fact that $\langle \theta({\bf k_2}) \theta({\bf k_1})\rangle  = \langle \langle \theta({\bf k_1})|\delta\rangle \langle \theta({\bf k_2})|\delta\rangle \rangle - \langle \alpha({\bf k_1}) \alpha({\bf k_2})\rangle $.

The quantities $\hat{f}$ and $\tilde{f}$ are readily extracted from simulations, for which we restrict the analysis of velocity and overdensity fields to large scales to avoid issues associated with the measurement of the velocity field in an unbiased way \citep{2009PhRvD..80d3504P, 2011MNRAS.410.2081J}.

 \begin{figure}
\begin{center}
\includegraphics[height=3in,width=3in]{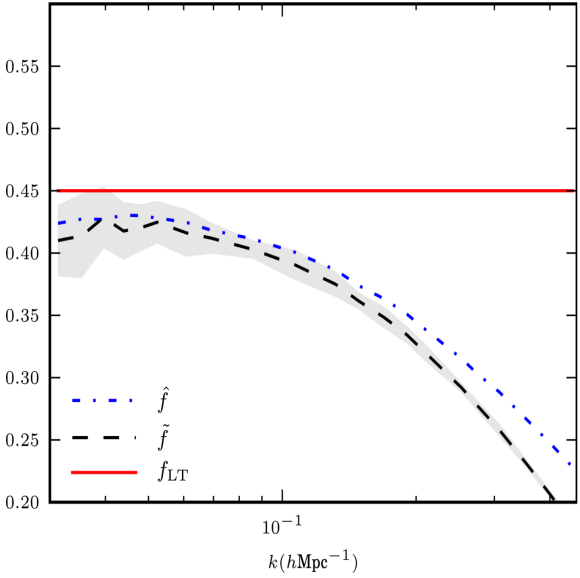}
\caption{The first moments of $\hat{f}$ and $\tilde{f}$, given in Eqs. (\ref{eq:fvariables1}) and (\ref{eq:fvariables2}), measured from the simulations at $z=0$ are shown as a blue dot dashed and black dashed line respectively. The linear theory growth rate is shown as a red solid line in this figure. The variance in $\tilde{f}$ measured from six realizations of the same cosmology is shown as the grey shaded region. \label{fig:7}}
\end{center}
\end{figure}

 As a side note, it is common to address the degree of stochasticity between two random functions $X$ and
$Y$ by measuring the cross correlation coefficient
$r = \langle X Y\rangle/\sqrt{\langle |X|^2\rangle \langle |Y|^2\rangle}$ as a function of scale.
This is a different notion of stochasticity to the one discussed in this paper, and relates to either a bias between the two fields at the level of the 
two point functions, $\langle |Y|^2\rangle = b_{2\rm{pt}} \langle |X|^2\rangle$  or a more specific local bias $Y = b_{\rm local}X$.
As pointed out by \citet{1999ApJ...520...24D},  the bias between the two point statistics follows from a local deterministic bias, and is the square of the local bias but the converse does not necessarily follow. In this case the cross correlation coefficient is a measure of $r = b_{2\rm{pt}}/b_{\rm local}$ and is not necessarily unity. Here the bias
$b_{\rm local}$ could represent the familiar bias between
the mass and halo/galaxy overdensity or we could view it as the
linear growth rate in the overdensity -- velocity divergence
relation in linear theory.
As pointed out by \citet{2004MNRAS.355..129S} the cross correlation coefficient can be close to unity despite fluctuations about
a local bias being large.

\subsection{Decomposition of two-point functions}

It is also instructive to decompose the two-point functions $\langle \theta(k_1) \delta(k_2)\rangle$ and $\langle \theta (k_1)\theta(k_2)\rangle$ into contributions coming from the nonlinear corrections and stochasticity in the $\theta$--$\delta$ relation.
The two-point function for $\alpha ({\bf k},t)$ in Eq. (\ref{eq:alpha}), decomposes as
\begin{equation*}
\langle \alpha({\bf k},t) \alpha^*({\bf k'},t)\rangle = \langle \langle \theta({\bf k},t) |\delta\rangle \langle \theta^*({\bf k'},t) |\delta\rangle \rangle -\langle \theta({\bf k},t) \theta^*({\bf k'},t) \rangle \, .
\end{equation*}
From this the two-point functions of interest -- the auto and cross power spectra between the conditional mean of $\langle \theta|\delta\rangle$ and $\delta$ -- can be expressed as
\begin{align}
\langle \theta_1 \delta_2^* \rangle &= \hat{f}_{12}\langle\delta_1\delta_2^*\rangle   + \langle\alpha_1 \delta_2^*\rangle  \label{eq:twopointfns1} \\ \nonumber
&=\langle \langle \theta_1|\delta\rangle \delta^*_2\rangle + \langle\alpha_1 \delta_2^*\rangle \\
 \langle \theta_1 \theta_2^*\rangle &=  \tilde{f}^2_{12} \langle\delta_1\delta_2^*\rangle + \langle \alpha_1 \alpha_2^*\rangle \label{eq:twopointfns2} \\ \nonumber
 &= \langle \langle \theta_1|\delta\rangle \langle \theta_2|\delta\rangle \rangle + \langle \alpha_1 \alpha_2^*\rangle.
\end{align}
Here we employ the short-hand notation $X_i$  for one-point quantities $X(\k_i)$ and $Y_{ij}$ for two-point quantities $Y(\k_i, \k_j)$. Note that $\hat{f}$ and $\tilde{f}$ are now evaluated as two-points functions. This separates out the nonlinear and stochastic components, as defined in Section \ref{sec:nonlinear} in a natural way, and emphasizes the different dependence on stochasticity for the auto-correlation and cross-correlation spectra. Also note that this approach is in contrast to previous studies \citep{Scoccimarro:2004tg, 2009MNRAS.393..297P, 2011MNRAS.410.2081J, 2012MNRAS.427L..25J,2012MNRAS.427.2537C, 2012MNRAS.425.2128J} which compare the ensemble averaged statistics $P_{\theta \delta} = \langle \theta_1 \delta_2^* \rangle $ and  $P_{\theta \theta} = \langle \theta_1 \theta_2^*\rangle$
with $P_{\delta \delta} = \langle \delta_1 \delta_2^*\rangle$ as a function of scale. In Section \ref{sec:results} we present the measurements of these two point functions and test the decomposition into nonlinear and stochastic components given in Eqs. (\ref{eq:twopointfns1}) and (\ref{eq:twopointfns2}).

\begin{figure*}
\begin{center}
\includegraphics[height=3.in,width=7.in]{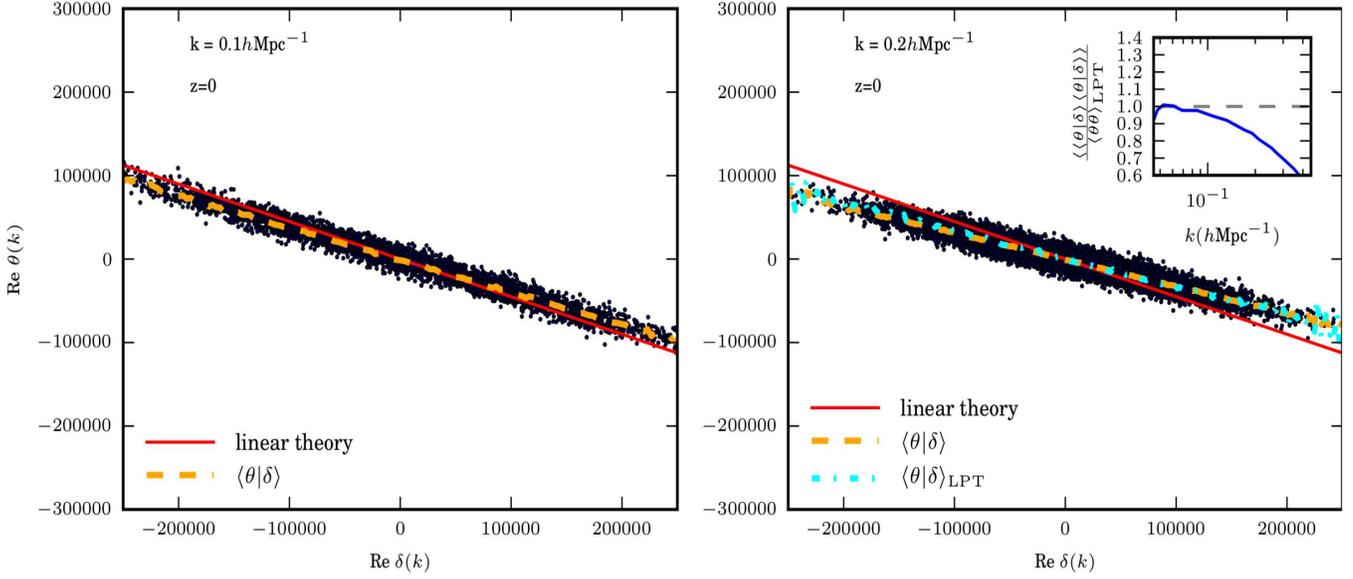}
\caption{The conditional expectation $\langle \theta({\bf {k}})|\delta\rangle$ (orange dashed line) at
$k = 0.1h$Mpc$^{-1}$ (left) and $k = 0.2h$Mpc$^{-1}$ (right) at $z=0$,
together with the linear theory relation between $\delta $ and $\theta$ (red line) and the real Fourier modes (black dots) measured from a $\Lambda$CDM simulation ($L_{\rm \tiny box} = 1500$Mpc$/h$) at $z=0$. In the right panel we show the conditional expectation $\langle \theta|\delta\rangle_{\tiny \rm LPT}$
from second order Lagrangian perturbation theory as a cyan dot dashed line. The inset panel shows the ratio of the two point function
$\langle\langle \theta|\delta\rangle\langle \theta|\delta\rangle\rangle/\langle \theta \theta\rangle_{\tiny \rm LPT}$ measured from the simulations at $z=0$ as a function of scale as a  blue solid line.
\label{fig:2}
}
\end{center}
\end{figure*}

\begin{figure*}
\begin{center}
\includegraphics[height=3.in,width=7.in]{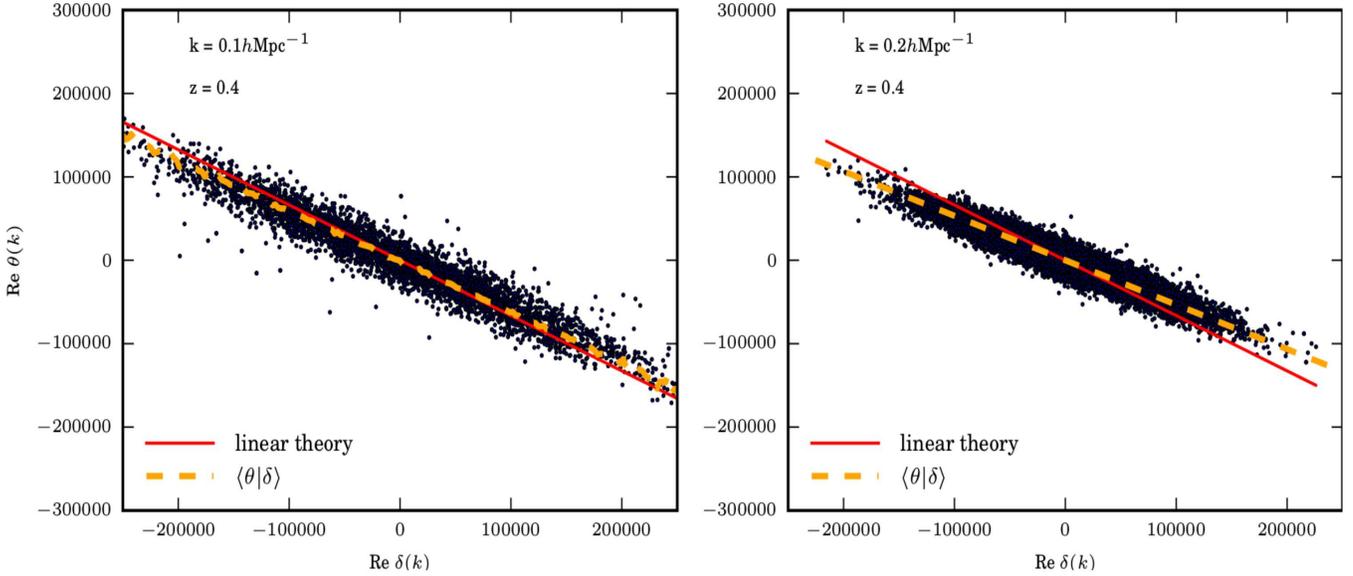}
\caption{The conditional expectation $\langle \theta({\bf {k}})|\delta\rangle$ (orange dashed line) at
$k = 0.1h$Mpc$^{-1}$ (left) and $k = 0.2h$Mpc$^{-1}$ (right) at $z=0.4$,
together with the linear theory relation between $\delta $ and $\theta$ (red line) and the real Fourier modes (black dots) measured from a $\Lambda$CDM simulation.
\label{fig:a0.7_1}
}
\end{center}
\end{figure*}

\section{Results}
 \label{sec:results}
 
We now provide a more detailed account of how the quantities introduced in the previous section behave in practice. The values of $ \hat{f}$ and $\tilde{f}$ are computed in Section \ref{sec:nl_f} at different scales, and compared with linear theory. The deviation of  $\theta$ from the conditional mean $\<\theta | \delta\>$ is addressed in Section \ref{sec:stoc}, both as a function of scale and redshift. We verify that the decomposition of the two point statistics into nonlinear and stochastic parts as defined in Section  \ref{sec:nonlinear} is reproduced within the simulation and we present the measured two point statistics in each case.
 
It turns out that halos display these features more dramatically than dark matter, and this is discussed in Section \ref{sec:halos}, where we measure $\langle \theta |\delta\rangle$ and the associated two point functions for halos with masses $M \le 5\times 10^{12}M_\odot h^{-1}$ from the MultiDark simulations.
An obvious question is: to what degree are these features reproduced by existing perturbative results, and do the decompositions presented simply correspond with a particular perturbative order? To this end, in Section \ref{sec:pt} we compare our results with standard perturbation theory to third order and second order Lagrangian Perturbation Theory (2LPT) predictions for the two point functions $\langle \theta \theta\rangle$ and $\langle \theta \delta\rangle$.

\subsection{Nonlinear  growth functions $\hat{f}$ and $\tilde{f}$}
 \label{sec:nl_f}

 The degree to which the moments given in  Eqs. (\ref{eq:fvariables1}) and (\ref{eq:fvariables2})  in Section \ref{sec:nonlinear} differ from $f_{LT}$ are a measure of the deviations from linearity, and provide effective non-linear growth rates. In Fig. \ref{fig:7} we plot these two moments, $\hat{f}$ and $\tilde{f}$, as a red solid (for $f_{LT}$), blue dot dashed and black dashed lines respectively, measured from the non-linear dark matter density field in the simulations  at $z=0$.
 
 Note that the two moments $\hat{f}$ and $\tilde{f}$ that are plotted are the average of six $N$-body simulations initialized with different realizations of the matter density field at early times. The variance of $\tilde{f}$ from these six simulations is shown as a grey shaded region.

 We find a notable difference between the three growth rates, and even on large scales, such as $k<0.1h$Mpc$^{-1}$, neither $\hat{f}$ nor $\tilde{f}$ correspond to the linear theory growth rate $f_{\tiny \rm LT}$. 
We find that the ratio of the two-point functions 
$\sqrt{\langle \theta \theta\rangle/\langle \delta \delta\rangle}$ 
and $\langle \delta \theta\rangle/\langle \delta \delta\rangle$ do converge 
to the linear theory result $f_{\tiny \rm LT}$ on much larger scales $k < 0.02 h$Mpc$^{-1}$ 
in agreement with previous work \citep{Scoccimarro:2004tg, 2009MNRAS.393..297P, 2011MNRAS.410.2081J, 2012MNRAS.427L..25J,2012MNRAS.427.2537C, 2012MNRAS.425.2128J}. Taking the decomposition of each of these two point functions given in Eqs. (\ref{eq:twopointfns1}) and (\ref{eq:twopointfns2}) into account this implies that on large scales the ratio of the stochastic two point functions $\langle \alpha \delta \rangle$ and $\langle \alpha \alpha \rangle$ to $\langle \delta \delta \rangle$ is at a minimum 10-15\% of  $f_{\tiny \rm LT}$ at $k < 0.1 h$Mpc$^{-1}$. We demonstrate that both of these decompositions are valid in Section \ref{sec:stoc}.

\subsection{The stochastic relation between $\theta$ and $\delta$.}
\label{sec:stoc}

In Fig \ref{fig:2} we plot the conditional expectation $\langle \theta |\delta \rangle$ as a orange dashed line. This is the average over six realizations, measured from the simulations by simply binning in $\rm{Re}\,\delta({\bf k})$ and finding the mean $\rm{Re}\,\theta({\bf k})$ at $z=0$ at the two scales $k=0.1h$Mpc$^{-1}$ (left panel) and $k=0.2h$Mpc$^{-1}$ (right panel), while the linear theory relation between $\delta $ and $\theta$ is plotted as a red line.
The real component of the Fourier modes measured from one simulation at each wavenumber are shown as black dots.
At each scale $k$ there is significant scatter between the Fourier modes measured from the simulations, and  $\langle \theta |\delta\rangle$ differs from linear perturbation theory predictions of $\theta = -f_{\tiny \rm LT}\delta$.
It is also clear from these two panels that the difference between $\langle \theta |\delta \rangle$ and  $-f_{\tiny \rm LT}\delta$
increases with increasing $k$ into the nonlinear  regime.

\begin{figure}
\begin{center}
\includegraphics[height=3in,width=3.in]{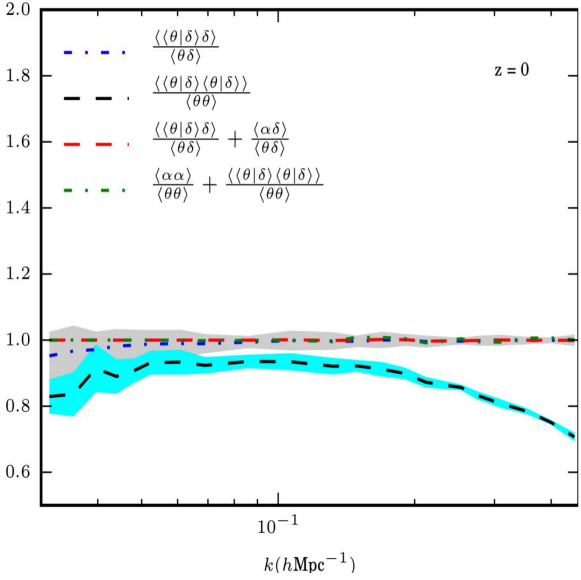}
\caption{The ratio of the two point functions in Eqs. (\ref{eq:twopointfns1}) and (\ref{eq:twopointfns2})
to $\langle \theta_1 \delta_2 \rangle $ and $\langle \theta_1 \theta_2 \rangle $ measured from the simulations at $z=0$ are shown as red dashed and green dot dashed lines respectively.
The ratios of the two point functions $\langle \langle\theta_1|\delta\rangle \delta_2 \rangle /\langle \theta_1 \delta_2 \rangle $ and $\langle \langle\theta_1|\delta\rangle \langle\theta_2|\delta\rangle \rangle /\langle \theta_1 \theta_2 \rangle $ are shown as a blue dot and black dashed line respectively. The shaded  cyan and grey regions show the variance of these ratios measured from six simulations.
 \label{fig:3}}
\end{center}
\end{figure}

\begin{figure}
\begin{center}
\includegraphics[height=4.7in,width=2.5in]{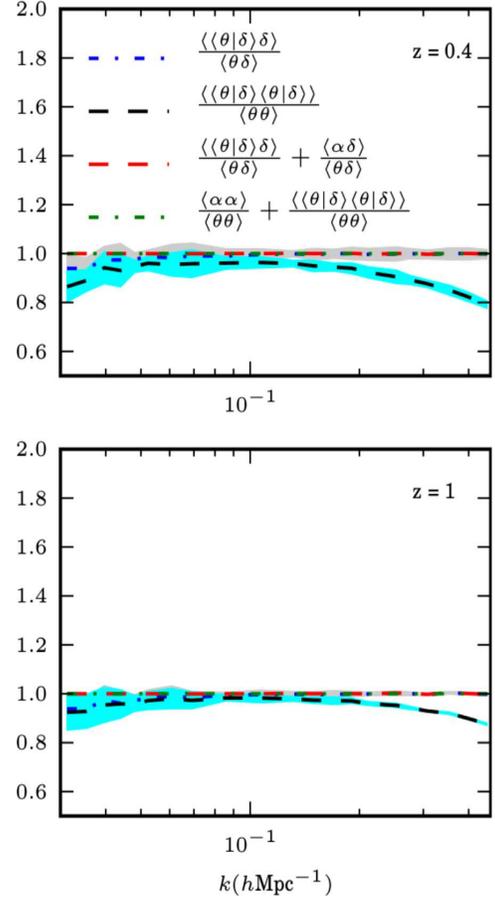}
\caption{The ratio of the two point functions in Eqs. (\ref{eq:twopointfns1}) and (\ref{eq:twopointfns2})
to $\langle \theta_1 \delta_2 \rangle $ and $\langle \theta_1 \delta_2 \rangle $ measured at at $z=0.4$ (top panel) and at $z=1$ (lower panel) are shown as red dashed and green dot dashed lines respectively. The ratios of the two point functions $\langle \langle\theta_1|\delta\rangle \delta_2 \rangle /\langle \theta_1 \delta_2 \rangle $ and $\langle \langle\theta_1|\delta\rangle \langle\theta_2|\delta\rangle \rangle /\langle \theta_1 \theta_2 \rangle $ are shown as a blue dot and black dashed line respectively. The shaded  cyan and grey regions show the variance of these ratios measured from six simulations.
 \label{fig:a0.7_2}}
\end{center}
\end{figure}

There are two notable effects which are evident from Fig \ref{fig:2}. Firstly, the nonlinearity we are describing with the conditional mean $\langle \theta |\delta \rangle$ manifests as an
approximate rotation about the linear theory prediction (orange dashed line in Fig. \ref{fig:2} compared to the solid red line) which is linear in $\delta$ but with a scale dependent coefficient i.e.
$\langle \theta |\delta \rangle \sim -f_{\tiny \rm LT}\delta + c(k) \delta$, where $c$ is an increasing function of scale. In Section \ref{sec:pt} we show that this functional dependence can be explained on large scales by second order Lagrangian perturbation theory. The second thing to note from these plots is that the stochastic scatter around $\langle \theta |\delta \rangle$ is nonzero and increases with increasing wavenumber $k$. At both scales we find that for $\delta > 0 (< 0)$ the mean relation $\langle \theta |\delta\rangle$ is larger (smaller) then the linear theory prediction, corresponding to an effective growth factor that is larger than linear theory.

The corresponding plot at $z=0.4$ is shown in Fig. \ref{fig:a0.7_1} for the same two scales. At higher redshifts we see the same trend with  $\langle \theta |\delta\rangle$ behaving as a rotation away from $-f_{\tiny \rm LT}\delta$ for linear theory. This difference increases with increasing wavenumber although this difference is smaller then at $z=0$ due to increased nonlinear growth at later redshifts as expected. We also note a reduction is the range of $\delta$ values at $z=0.4$ compared to $z=0$. An identical number of modes have been used at each scale and redshift.

\begin{figure*}
\begin{center}
\includegraphics[height=3.in,width=7.in]{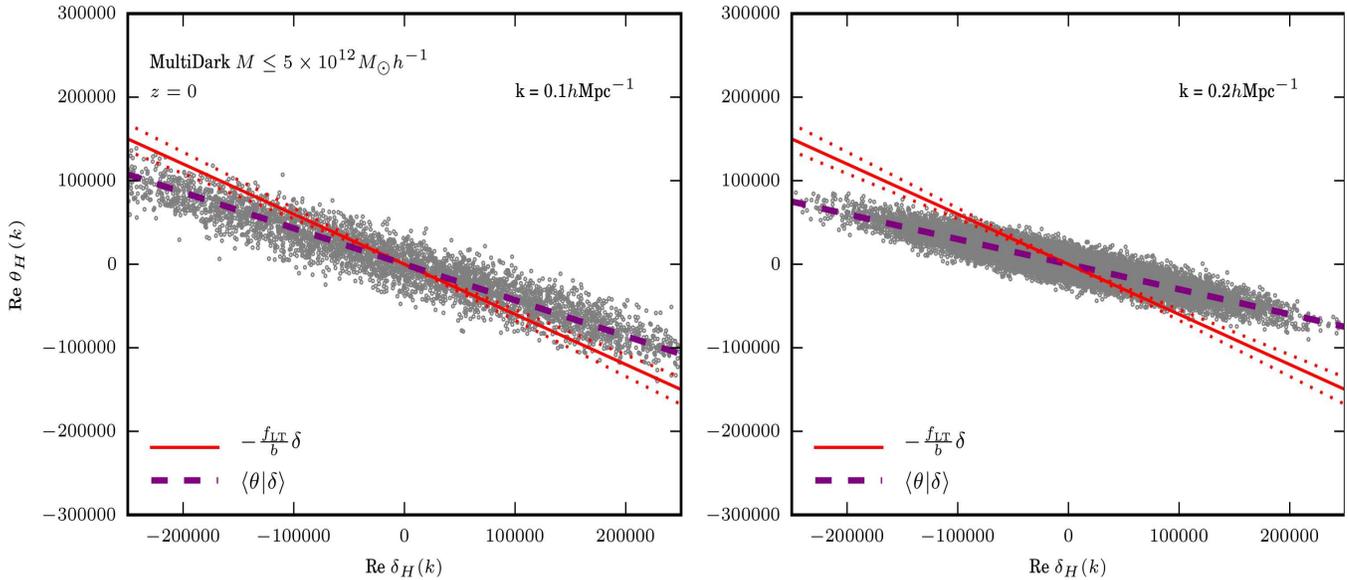}
\caption{The conditional expectation $\langle \theta({\bf {k}})|\delta\rangle$ (purple dashed line) measured using halos with $M < 5\times 10^{12}M_\odot h^{-1}$ in the MultiDark simulation
 at
$k = 0.1h$Mpc$^{-1}$ (left) and $k = 0.2h$Mpc$^{-1}$ (right) at $z=0$.
The linear theory relation $\theta_H = -f_{\tiny \rm LT}/b \delta_H$, where $b$ is the linear bias on large scales is shown as a red  solid line.
The red dotted lines either side of the linear theory prediction represent a $\pm 10$\% error in the linear bias.
The real Fourier modes measured using this halo catalogue at each wavenumber are shown as grey dots.
\label{fig:MD}
}
\end{center}
\end{figure*}

The decomposition of the two point functions $\langle \theta \delta \rangle$ and $\langle \theta \theta \rangle$ into nonlinear and stochastic parts, as in Eqs. (\ref{eq:twopointfns1}) and (\ref{eq:twopointfns2}), is readily verified numerically. In Fig. \ref{fig:3} we plot the ratios of the RHS of Eqs. (\ref{eq:twopointfns1}) and (\ref{eq:twopointfns2}) to $\langle \theta_1 \delta_2 \rangle $ and $\langle \theta_1 \delta_2 \rangle $ as red dashed and green dot dashed lines respectively. We find these ratios are unity which verifies the decompositions in Eqs. (\ref{eq:twopointfns1}) and (\ref{eq:twopointfns2}) from the simulations. This result is non-trivial as all of the two point statistics have been measured independently from the simulations i.e. $\langle \langle \theta|\delta\rangle \delta\rangle$ is an ensemble average over the mean $\theta$ given $\delta$ (orange dashed line in  Fig. \ref{fig:2}) times $\delta$. This is in contrast to $\langle \theta \delta\rangle$ which is the ensemble average over each $\theta$ and $\delta$ (black dots in Fig. \ref{fig:2}).

In Fig \ref{fig:3} we also plot the ratios $\langle \langle \theta|\delta\rangle \langle \theta|\delta\rangle\rangle/\langle \theta \theta \rangle$ and $\langle \langle \theta|\delta\rangle\delta \rangle/\langle \theta \delta \rangle$ measured from the simulations at $z=0$ as black dashed and blue dot dashed lines respectively. The shaded regions in this plot represent the variance amongst six realizations of the same cosmology.
We find that the stochastic components contribute $\sim 10$\% to the two point function $\langle \theta_1 \theta_2\rangle$
while it contributes about a 1\% to  $\langle \theta_1 \delta_2\rangle$ at $k<0.2h$Mpc$^{-1}$. The stochastic component of the velocity divergence auto power increases to approximately 25\% by $k = 0.45 h$Mpc$^{-1}$.
In 
the upper and lower panels of
Fig. \ref{fig:a0.7_2} we show similar power spectra ratios to those in Fig. \ref{fig:3} at $z=0.4$ and 
$z=1$ respectively. We find that the stochastic
component of the velocity divergence power spectrum is slightly reduced at higher redshifts as there is less nonlinear growth present at earlier times which would induce a larger variation in
$\theta$ from $\langle \theta |\delta\rangle$.

\begin{figure}
\begin{center}
\includegraphics[height=3in,width=3.in]{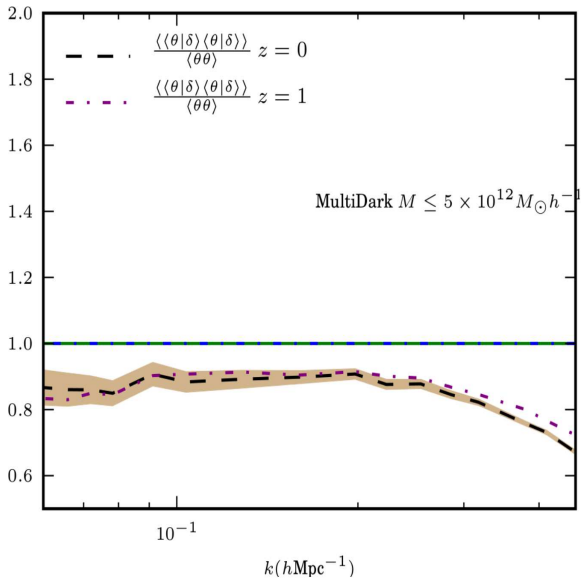}
\caption{The ratio of the two point function
 $\langle \langle\theta_1|\delta\rangle \langle\theta_2|\delta\rangle \rangle /\langle \theta_1 \theta_2 \rangle $
 measured at $z=0$ and $z=1$ using halos with
 $M < 5\times 10^{12}M_\odot h^{-1}$ from the MultiDark simulation are shown as
 black dashed and blue dot dashed lines respectively. The shaded tan region represents the error on the measured power spectra given in Eq. (\ref{eq:pkerror}) in Section \ref{sec:sims}.
 \label{fig:halos_2}}
\end{center}
\end{figure}

\subsection{Behaviour of $\langle \theta|\delta \rangle$ and $\alpha$ for dark matter halos}
\label{sec:halos}

In Fig. \ref{fig:MD} we show the conditional expectation $\langle \theta({\bf {k}})|\delta\rangle$ 
as a purple dashed line, measured 
at $k = 0.1h$Mpc$^{-1}$ (left panel) and $k = 0.2h$Mpc$^{-1}$ (right panel)
from the MultiDark simulations using halos with
masses $M < 5\times 10^{12}M_\odot h^{-1}$.
The real Fourier modes Re$\theta_H(k)$ and Re$\delta_H(k)$ are shown as grey dots in both panels. We plot the linear theory prediction $\theta_H = -f_{\tiny \rm LT}/b \delta_H$, where $b$ is the linear bias on large scales as a red  solid line. The red dotted lines either side of the linear theory prediction represent a $\pm 10$\% error in the linear bias. For this halo sample we find that the linear bias is $b \sim 0.81 \pm 0.09$ and is reasonably linear on scales $k \le 0.2 h$Mpc$^{-1}$ \citep[see also e.g.][]{2015MNRAS.446..793J}.

It is clear that there is significant scatter about the mean  $\langle \theta({\bf {k}})|\delta\rangle$ and that this conditional expectation differs from the linear theory prediction by an approximate rotation.
If we compare these results with Fig. \ref{fig:2} in Section \ref{sec:stoc} for the dark matter we see that at the same redshift, the deviation of $\langle \theta({\bf {k}})|\delta\rangle$ from the linear theory prediction and the scatter about the conditional mean given by $\alpha$, is much larger for the halo sample then for the dark matter at both $k$ scales. Note these two simulations have slightly different cosmologies, for example $\Omega_m = 0.24$ ($0.27$) in the dark matter (Multidark) simulations, which may account for some of these  differences.

In Fig. \ref{fig:halos_2} we plot the ratio of the two point function
 $\langle \langle\theta_1|\delta\rangle \langle\theta_2|\delta\rangle \rangle /\langle \theta_1 \theta_2 \rangle $
 measured at $z=0$ and $z=1$ for the same halo sample as black dashed and 
 purple dot dashed lines respectively.  The shaded tan region represents the error on the measured power spectra given in Eq. (\ref{eq:pkerror}) in Section \ref{sec:sims}.  We have also verified that the decomposition of the two point functions into nonlinear and stochastic parts, as given in Eqs. (\ref{eq:twopointfns1}) and (\ref{eq:twopointfns2}), holds for the halo two point functions. We have omitted this from Fig.  \ref{fig:halos_2} for clarity.  Therefore any deviation from unity in this figure indicates the magnitude of the stochastic component.
 We find that the stochastic component of the two point function $\langle \theta_1 \theta_2 \rangle $ is significant and approximately a constant fraction ($\sim$15\%) at $k<0.25 h$Mpc$^{-1}$ at both $z=0$ and $z=1$. The differences between the halo sample and the dark matter distribution, in how the conditional mean deviates from the linear theory predictions and the scatter
 around that mean as a function of wavenumber, cannot be only due to a difference in cosmological parameters. As shown in Fig. \ref{fig:halos_2} we find that the velocity divergence
 auto power spectrum has a larger stochastic component in the halo sample compared to the dark matter at both redshifts.

\begin{figure}
\begin{center}
\includegraphics[height=3in,width=3in]{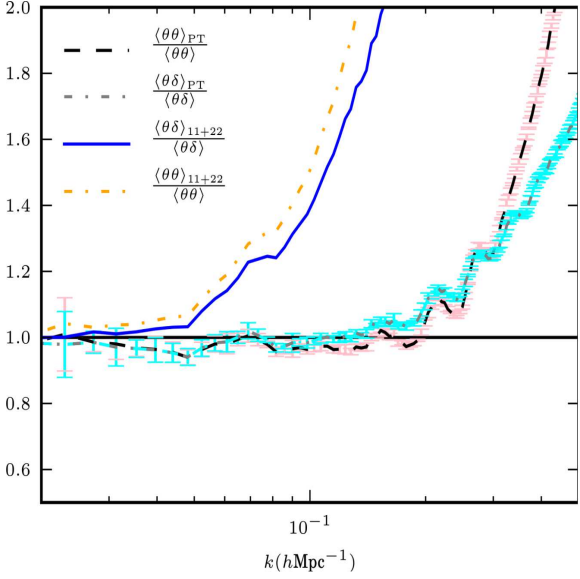}
\caption{The ratio of the one loop perturbation theory prediction for the
velocity divergence cross and auto power spectra (Eq. (\ref{eq:spt})) to $\langle \theta \delta\rangle$ and $\langle \theta \theta\rangle$ measured from the dark matter only simulations
are shown a grey dot dashed and black dashed line
respectively. The pink and cyan
error bars show the variance in these ratios from six simulations with different
realisations of the initial density field. Similar ratios of
 $P_{\rm \tiny 11} + P_{22}$ calculated from one loop perturbation theory for the both cross and auto power spectrum are
shown as blue solid and orange dot dashed lines as given in the legend.
\label{fig:4}}
\end{center}
\end{figure}

\subsection{The relation to standard and Lagrangian Perturbation Theory}
\label{sec:pt}

In this section we connect the formalism presented in Section \ref{sec:nonlinear} to perturbation theory methods.
First we consider standard perturbation theory predictions for both the velocity divergence auto and cross power spectra, $\langle \theta \theta\rangle$ and
$\langle \theta \delta\rangle$, computed up to third order \citep[see e.g.][for a review]{2002PhR...367....1B}.
The nonlinear velocity divergence auto $P({\bf k})$ computed from third order perturbation theory is given by
\begin{equation}
\langle \theta \theta\rangle_{\tiny \rm PT} ({\bf k})= P({\bf k}) + P_{22}({\bf k}) + 2P_{13}({\bf k}) \label{eq:spt}\,
\end{equation}
where $P({\bf k})$ denotes the linear power spectrum and the scale dependent functions $P_{22}$ and $P_{13}$
are given by
\begin{align}
P_{22}({\bf k})  &= 6P({\bf k}) \int {\rm d}^3qG_3({\bf k}, {\bf q}) P({\bf q}) \\
P_{13}({\bf k}) &= \int {\rm d}^3q [G_2( {\bf k}-{\bf q} ,{\bf q})]^2 P({\bf k} - {\bf q}) P({\bf q}) \, ,
\end{align}
where the kernel $G_2$ is given by
\begin{align}
G_2( {\bf k} ,{\bf q}) &= \frac{\mu_2}{2} + \frac{1}{2}\hat{k}\cdot \hat{q} \left( \frac{k}{q} + \frac{q}{k}\right) \nonumber \\
& + \frac{4}{7}\left(\hat{k}_i \hat{k}_j  - \frac{1}{3}\delta_{ij} \right) \left( \hat{q}_i \hat{q}_j  - \frac{1}{3}\delta_{ij} \right) \label{eq:kernel},
\end{align}
where $\mu_2 = 26/21$
and the angle averaged $G_3$  kernel is given in e.g. Eq. (69) in \citep{Scoccimarro:2004tg}.
A similar expression for the velocity divergence cross power spectrum to third order can also be found in
\citep{Scoccimarro:2004tg}.

In order to compare the formalism in this paper, which decomposes the velocity divergence two point statistics into nonlinear and stochastic elements as given in Section \ref{sec:nonlinear}, with perturbation theory methods we simply calculate the  individual power spectra in Eq. (\ref{eq:spt}) and compare them with the measure two point velocity divergence statistics.
In Fig \ref{fig:4} we show the ratio of the one loop perturbation theory predictions for the velocity divergence cross and auto power spectra to $\langle \theta \delta\rangle$ and $\langle \theta \theta\rangle$, measured from the dark matter only simulations at $z=0$,
as a grey dot dashed and black dashed line respectively. The pink and cyan error bars show the variance in these ratios from six
simulations with different realizations of the initial density field. The ratios of $P_{\rm \tiny 11} + P_{22}$ calculated from 1 loop perturbation theory for the both cross and auto power spectrum to $\langle \theta \delta\rangle$ and $\langle \theta \theta\rangle$ are
shown as blue solid and orange dot dashed line respectively.

By comparing Figs. \ref{fig:3} and \ref{fig:4}  we can see that at the level of 3rd order perturbation theory that the standard perturbation theory prediction and the formalism in this paper deviate substantially and no simple identification can be made.  Even on large scales, $k < 0.05h$Mpc$^{-1}$, where the perturbation theory predictions match the measured power spectra from the simulations to $\sim 5$\% we cannot simply relate the mode coupling terms $P_{13}$, which are negative, to the
stochastic power spectra $\langle \alpha\alpha \rangle$ and $\langle \alpha\delta \rangle$.

Next we consider second order Lagrangian perturbation theory (2LPT) predictions for the $\theta -\delta $ relation
\citep[][]{1993ApJ...405L..47G, 1995A&A...296..575B,1996dmu..conf..565B,1995A&A...294..345M, 2012MNRAS.425.2422K}. Lagrangian perturbation theory represents a alternative framework to the Eulerian approach, and the non-linear analysis is based around the trajectories of individual fluid elements. Of central importance is the displacement field $\Psi({\bf q})$, which provides a mapping from initial Lagrangian coordinates ${\bf q}$ to final Eulerian coordinates ${\bf x}$  given by ${\bf x}(\tau)  = {\bf q} + \Psi({\bf q},\tau)$ \citep[see e.g.][ for a review]{1996dmu..conf..565B, 2002PhR...367....1B}.
The linear solution for the equations of motion coincide with the Zel'dovich approximation,
\begin{equation}
\nabla_{\bf q}\cdot \Psi^{(1)} = -D(\tau)\delta^{(1)}({\bf q}) \, ,
\end{equation}
where $\delta^{(1)}({\bf q})$ is the linear density field and $D$ is the linear growth factor normalized to unity at $z=0$.
In contrast, the second order correction to the displacement field  \citep[see e.g.][]{1995A&A...294..345M} takes into account tidal gravitational effects as
\begin{equation}
\nabla_{\bf q}\cdot \Psi^{(2)} = \frac{1}{2}D_2\sum_{i \ne j} \left( \Psi^{(1)}_{i,i}\Psi^{(1)}_{j,j} - [\Psi^{(1)}_{i,j}]^2\right) \, ,
\end{equation}
where $\Psi^{(1)}_{i,j} = \partial \Psi_i/\partial q_j$ and $D_2$ is the second order growth factor given by $D_2 \approx -3/7 D^2 \Omega_m^{1/143}$.
The Lagrangian potentials $\phi^{(1)}$ and $\phi^{(2)}$ are defined such that $\nabla^2\phi^{(1)} ({\bf q})= \delta^{(1)}({\bf q})$ and $\nabla^2\phi^{(2)} ({\bf q})= \delta^{(2)}({\bf q})$. The 2LPT expressions for the position become
\begin{align}
x({\bf q}) &= {\bf q} - D\nabla_{q}\phi^{(1)}  + D_2  \nabla_{q}\phi^{(2)} \, ,
\end{align}
while the dimensionless velocity divergence is given by
\begin{equation}
\theta = -D f_{\tiny \rm LT} \nabla^2_{q}\phi^{(1)}  + D_2  f_2 \nabla^2_{q}\phi^{(2)} \label{eq:theta_lpt}\, ,
\end{equation}
where $f_{\rm \tiny LT}$ is the linear theory growth rate and $f_2  = {\rm d ln}D_2/{\rm d ln}a$ is the logarithmic derivative of the second-order growth factor, $f_2 \approx 2\Omega_m^{6/11}$. The gradient terms are given by
\begin{align}
\nabla^2_{q}\phi^{(1)} &= \delta^{(1)}({\bf q}) \label{eq:lpt_1}\\
\nabla^2_{q}\phi^{(2)} &= \sum_{i > j} \left( \phi^{(1)}_{,ii}\phi^{(1)}_{,jj} - [\phi^{(1)}_{,ij}]^2\right) \label{eq:lpt_2} \, ,
\end{align}
where $\phi^{(1)}_{,ii} = \partial^2 \phi/\partial q_i \partial q_j$. These equations imply that given a robust estimate of the linear overdensity $\delta^{(1)}$ then we can obtain a corresponding non-linear velocity divergence $\theta$ from 2LPT.

To compare the conditional mean  $\langle \theta|\delta\rangle$ measured from the simulations in Fourier space with $\langle \theta | \delta \rangle_{\tiny \rm LPT}$, where $\theta$ is the Fourier transform of the 2LPT prediction in Eq. (\ref{eq:theta_lpt}),
 we estimate the linear density field $\delta^{(1)}$ from the nonlinear matter field $\delta_m$ in real space as given by \citet{2009ApJ...698L..90N}, $\delta^{(1)} = {\rm log}(1+\delta_m) - \langle{\rm log}(1+\delta_m) \rangle$. The quantity $\phi^{(2)}(\k)$ can be obtained by Fourier transforming $\phi^{(1)}(\k)_{,ij}$ into real space, computing the sum and then transforming back. Alternatively one can Fourier transform Eq. (\ref{eq:lpt_2}) directly, and then obtain the total 2LPT prediction for $\theta({\bf k})$.

In  the right panel of Fig. \ref{fig:2} we plot the conditional expectation
$\langle \theta|\delta\rangle_{\tiny \rm LPT}$, which has been evaluated by the same method as described in Section \ref{sec:stoc} using  the 2LPT prediction for $\theta({\bf k})$, as a cyan dot dashed line.
From this figure it seems that the rotation of the conditional mean away from the linear theory prediction is captured well by 2LPT at $k=0.2 h$Mpc$^{-1}$. In the inset panel of Fig. \ref{fig:2} we also show the ratio of the two-point function
$\langle\langle \theta|\delta\rangle\langle \theta|\delta\rangle\rangle/\langle \theta \theta\rangle_{\tiny \rm LPT}$ measured from the simulations at $z=0$ as a function of scale as a blue solid line.
On scales $k<0.1 h$Mpc$^{-1}$ we can see that the ratio is very close to unity indicating that the nonlinear effects in the formalism of this paper can be described by 2LPT which incorporate the effects of tidal gravitational fields on large scales.
 Recall that the full two point function $\langle \theta \theta \rangle$ can be written as sum of nonlinear and stochastic components. Our results indicate that the stochastic component, which is approximately 15\% of the velocity divergence auto power spectrum at $z=0$ and $k=0.1 h$Mpc$^{-1}$
is not described by the predictions of 2LPT. At smaller scales $k>0.1 h$Mpc$^{-1}$ we find that the $\theta$ -- $\delta$ relation is still well-described by the combined action of a non-linear rotation together with stochastic spread, however the predictions of 2LPT no longer adequately describe the regime.

\section{Redshift space distortions}\label{sec:rsd}

We begin in Section \ref{sec:rsd_1} by briefly reviewing the theory of redshift space distortions and models that depend on the linear growth rate $f_{\tiny \rm LT}$ and which are currently in use. In Section \ref{sec:rsd_2} we highlight the problems associated with having a well defined notion of the linear growth rate in a redshift space distortion model in the presence of a nonlinear and stochastic $\theta - \delta$ relation.

\subsection{Redshift space distortion models }
\label{sec:rsd_1}

Inhomogeneous structure in the Universe induces peculiar motions which distort the clustering
pattern measured in redshift space on all scales. This effect must be taken into account when analyzing three dimensional datasets that use
redshift to estimate the radial coordinate.
Redshift space effects alter the appearance of the clustering
of matter, and together with nonlinear evolution and bias, lead the measured
power spectrum to depart from  the simple predictions of linear perturbation theory.
The comoving distance to a galaxy, $\vec{s}$,  differs from its true distance, $\vec{x}$, due to its peculiar velocity, $\vec{v}(\vec{x})$
(i.e. an additional velocity to the Hubble flow). The mapping from redshift space to real space is given by
\begin{equation}
\vec{s} = \vec{x} + u_z \hat{z},
\end{equation}
where $u_z = \vec{v}\cdot \hat{z}/(aH)$ and $H(a)$ is the Hubble parameter. This assumes that the distortions take place along the line of sight, denoted by $\hat{z}$, and is commonly referred to as the plane parallel approximation.

On small scales, randomised velocities associated with the motion of galaxies inside virialised structures reduce the power.
The dense central regions of galaxy clusters appear elongated along the line of sight in redshift space, which produces the  \lq fingers
of God\rq\
effect in redshift survey plots.
For growing perturbations on large scales, the overall effect of redshift space distortions is to enhance the clustering amplitude.
Any difference in the velocity field due to mass flowing from underdense regions to high density regions will alter the volume element, causing
an enhancement of the apparent density contrast in redshift space, $\delta_s(\vec{k})$, compared to that in real space, $\delta_r(\vec{k})$ \citep[see][for a review of redshift space distortions]{1998ASSL..231..185H}.

Assuming the line of sight component of the peculiar velocity
is along the $z$-axis, the power spectrum in redshift space is given by \citep{Scoccimarro:2004tg}
\begin{align}
\delta_D(\vec{k}) + P_s(\vec{k}) =\int \!\! \frac{{\rm d}^3r}{(2\pi)^3} e^{-i\vec{k}\cdot\vec{r}} \langle e^{ik_zV}[1+\delta_g(\vec{x})][1+\delta_g(\vec{x}')]\rangle
\end{align}
where $\delta_g = b \delta$ is the galaxy overdensity which we shall assume is related by a linear bias, $b$ to the matter overdensity, $V = u_z(\vec{x}) - u_z(\vec{x}')$ and $\vec{r} = \vec{x} - \vec{x'}$. We are also assuming that there is no velocity bias between the dark matter and galaxies for simplicity.

Decomposing the vector field into curl and divergence free parts, and assuming an irrotational velocity field, we can re-write
$k_z u_z =  -(k_z^2/k^2 )\theta(k)  = -\mu^2 \theta(k)$ where $\theta(k)$ is the Fourier transform of the velocity divergence defined in Eq. (\ref{eqn.linearcontinuity}).
Expanding the exponential term and only keeping terms up to second order in the variables $\delta$ and $\theta$, the power spectrum in redshift space $P_s$ becomes
\begin{align}\label{eq:nl_rsd}
\delta_D(\vec{k} - \vec{k'})P_s(\vec{k}) &=b^2\langle \delta(\vec{k})\delta^*(\vec{k'}) \rangle -2\mu^2 b \langle \theta(\vec{k})\delta^*(\vec{k}') \rangle \nonumber \\
&+ \mu^4\langle \theta(\vec{k})\theta^*(\vec{k}') \rangle.
\end{align}
If we assume the linear continuity equation holds we can re-write this as
\begin{align}
\nonumber
\delta_D(\vec{k} - \vec{k'})P_s(\vec{k}) &=\langle \delta(\vec{k})\delta^*(\vec{k'})\rangle [b^2  -2bf_{\rm \tiny LT}\mu^2 + f_{\rm \tiny LT}^2 \mu^4] \\ \nonumber
&=  \delta_D(\vec{k}-\vec{k'})P(k)[b^2  - 2bf_{\rm \tiny LT}\mu^2 + f_{\rm \tiny LT}^2 \mu^4]  \label{eq:kaiser}\\
\end{align}
which is the \citet{1987MNRAS.227....1K} formula for the power spectrum in redshift space in terms of the linear growth rate $ f_{\rm \tiny LT}$, the linear bias $b$ and the power spectrum $P(k)$.

Commonly used models for the redshift space power spectrum extend the Kaiser formula by assuming that the velocity and density fields are uncorrelated and that the joint
probability distribution factorizes as  $\mathcal{P}(\delta,\theta) = \mathcal{P}(\theta)\mathcal{P}(\delta)$.
Examples include multiplying Eq. (\ref{eq:kaiser}) by a factor which
attempts to take into account small scale effects, invoking either a
Gaussian or exponential distribution
of peculiar velocities.
A popular phenomenological example of this which incorporates the damping effect of velocity dispersion on small scales is the so-called \lq dispersion model\rq \ \citep{1994MNRAS.267.1020P},
\begin{equation}
P^s(k,\mu) =  P_g(k) (1+\beta \mu^2)^2 \frac{1}{(1 + k^2 \mu^2 \sigma_p^2/2)} 
\label{eq:pd}\, ,
\end{equation}
where $P_g$ is the galaxy power spectrum, $\sigma_p$ is the pairwise velocity dispersion along the line of sight, which is treated as a parameter to be fitted to the data and $\beta = f_{\tiny \rm LT}/b$.

The linear model for the redshift space power spectrum can be extended by keeping
the nonlinear velocity power spectra terms in Eq. (\ref{eq:nl_rsd}).
For example \citet{Scoccimarro:2004tg} proposed the following model for the redshift space power spectrum in terms of $P_{\delta \delta}$, the nonlinear matter power spectrum,
\begin{align}\label{SM}
P^s(k,\mu)&=  \left( P_{\delta \delta}(k) + 2 \mu^2 P_{\delta \theta}(k) + \mu^4P_{\theta \theta}(k)\right)
\times e^{-( k \mu \sigma_v )^2} ,
\end{align}
where  $P_{\theta \theta} = \langle \theta \theta\rangle $,  $P_{\delta \theta}= \langle \delta \theta \rangle$ and $\sigma_v$ is the 1D linear velocity dispersion given by
\begin{equation}
 \sigma^2_v = \frac{1}{3}\int\frac{P_{\theta \theta}(k)}{k^2} {\rm d}^3k.
\end{equation}
In linear theory, $P_{\theta \theta}$ and $P_{\delta \theta}$ take the same form as $P_{\delta \delta}$ and depart from this at different scales.
Using  a simulation  with 512$^3$ particles in a box of length $479 h^{-1}$Mpc, 
\citet{Scoccimarro:2004tg} showed that this simple ansatz  for $P_s(k,\mu)$ was an improvement over the Kaiser formula when comparing to
the results of $N$-body simulations in a $\Lambda$CDM
cosmology. Clearly the inclusion of these nonlinear velocity divergence terms gives rise to an improved model of redshift space distortions in the nonlinear regime.

In  nonlinear models for the power spectrum in redshift space there is a degeneracy between the
nonlinear bias, the difference between the clustering of dark matter and halos or galaxies, and the
scale dependent damping due to velocity distortions on small scales.
This degeneracy will complicate any measurement of the growth rate using redshift space clustering information on small scales.
In this work we have restricted our analysis of the $\theta - \delta$ relation to large scales for the halo sample where the approximation of a linear bias is valid. Note also that nonlinearities in the bias between the halos and dark matter field affect the $\mu^2$ component but not the $\mu^4$ coefficient if there is no velocity bias present.

\subsection{Modeling redshift space distortions with a nonlinear stochastic $\theta - \delta$ relation}
\label{sec:rsd_2}

Firstly, the expansion in Eq. (\ref{eq:nl_rsd}) does not assume that $\theta$ and $\delta$ are uncorrelated ($\mathcal{P}(\delta,\theta) = \mathcal{P}(\theta)\mathcal{P}(\delta)$) but
instead only retains terms which are second order in  $\theta$ and $\delta$. We can rewrite Eq. (\ref{eq:nl_rsd}) in terms of the main formalism in this paper which describes a nonlinear, stochastic relation between $\theta$ and $\delta$. Using Eqs. (\ref{eq:twopointfns1}) and (\ref{eq:twopointfns2}) with the adapted notation
$\langle \theta(\vec{k}) \theta(\vec{k'})\rangle =  \langle \theta_1 \theta_2 \rangle$ etc.
we can write
\begin{align}
\label{eq:stoc_rsd}
\delta_D(\vec{k} - \vec{k'})P_s(\vec{k}) =&  \, b^2\langle \delta(\vec{k})\delta^*(\vec{k'}) \rangle\\ \nonumber
&  -2b\mu^2 [ \langle \langle \theta_1|\delta\rangle\delta_2 \rangle + \langle\alpha_1 \delta_2\rangle]  \\ \nonumber
&+ \mu^4[ \langle \alpha_1 \alpha_2\rangle + \langle \langle \theta_1|\delta\rangle \langle \theta_2|\delta\rangle\rangle].
\end{align}
There are a small number of papers that have used perturbation theory to find an analytic formula for the conditional mean $\langle \theta | \delta \rangle$
\citep[see e.g.][]{1998astro.ph..6087C,1999MNRAS.309..543B}. Guided by the results in Section \ref{sec:nonlinear} where   $\langle \theta | \delta \rangle$  appears as a rotation from the linear
perturbation theory prediction which increased with increasing wavenumber $k$, we consider the following simple expression for
$\langle \theta({\bf k}) | \delta \rangle = -f_{\tiny \rm LT}\delta({\bf k}) + c({\bf k})\delta({\bf k})$. Putting this into Eq. (\ref{eq:stoc_rsd}) we obtain the expression
\begin{align}
\delta_D(\vec{k} - \vec{k'})P_s(\vec{k}) =& 
\langle \delta_1\delta^*_2 \rangle \left (  b^2 - 2b\mu^2[-f_{\tiny \rm LT} + c({\bf k}) + \frac{\langle\alpha_1 \delta_2\rangle}{\langle \delta_1\delta^*_2 \rangle} ] \right .\nonumber \\
&\left . + \mu^4[ ( f_{\tiny \rm LT} - c({\bf k}))^2+ \frac{\langle \alpha_1 \alpha_2\rangle}{\langle \delta_1\delta^*_2 \rangle}] \right ).
\end{align}
A key point that this highlights is that the coefficients in front of the $\mu^2$ and $\mu^4$ terms no longer have a simple relation. The receive different contributions from non-linearity and stochasticity, and cannot be simply written as $f_{\rm \tiny NL} \mu^2 + f_{\rm \tiny NL}^2 \mu^4$.
If the relation between $\theta $ and $\delta$ is deterministic ($ \langle \alpha_1 \delta_2\rangle = 0$ and $\langle \alpha_1 \alpha_2\rangle =0$) then,
as shown in Section \ref{sec:pt}, second order Lagrangian perturbation theory provides a good description of the nonlinear rotation of the conditional mean $\langle \theta | \delta \rangle$ away from the linear perturbation theory predictions at $k<0.1 h$Mpc$^{-1}$.
The stochastic components $ \langle\alpha_1 \delta_2\rangle$ and $\langle \alpha_1 \alpha_2\rangle$ are nonzero at $z=0, 0.4$ and $z=0$ on large scales, as can be seen from Figs. \ref{fig:3} and \ref{fig:a0.7_2}, and comprise approximately 10\% of the velocity divergence auto power spectrum on large scales $k<0.1 h$Mpc$^{-1}$.

It is common practice to try to extract a measurement of the linear growth rate, $f_{\rm \tiny NL}$,
using the $\mu^2$ and $\mu^4$ dependence of the measured galaxy power spectrum in redshift space,
 and either  the model in Eq. (\ref{eq:pd}) or models which include the velocity divergence auto and cross power spectra.
If however there is a nonlinear and stochastic relation between $\theta$ and $\delta$ then the correspondence between the coefficients of $\mu^2$ and  $\mu^4$, and $f_{\rm \tiny NL}$ becomes more complex. 

Ideally a perturbative expansion which captures all the nonlinearities in Eqs. (\ref{eq:nl1}) and (\ref{eq:nl2}) would give an accurate prediction for the velocity divergence
two and higher point statistics and their correlations with the matter overdensity. This would include the stochastic terms in the formalism in this paper which are produced by nonlinear effects. Without this exact expansion, it is 
not straightforward to make an explicit connection between
the quantity of interest, $f_{\rm \tiny NL}$, and parameters in current phenomenological models for two point clustering statistics in redshift space, which either assume
that $\theta$ and $\delta$ are related by a linear, deterministic  relation, or are based on perturbation theory expressions to a given order for the $\theta -\delta$ relation.

\section{Summary \& conclusions}
\label{sec:conc}

Up and coming galaxy redshift surveys aim to measure the linear growth rate to an accuracy of $\sim 1$\%. This growth rate is commonly obtained from a deterministic relation between the velocity divergence and the matter overdensity fields that follows from linear theory.
Here we have explored a formalism that defines both a nonlinear and a stochastic relation between the velocity divergence and overdensity field, $\theta = \nabla \cdot v({\bf x},t)/aH$ and $\delta$, which is based on an extension of linear theory to a relation in terms of the conditional mean $\langle \theta|\delta\rangle$, together with fluctuations of $\theta$ around this non-linear relation.

Using N-body simulations of
dark matter particles that follow the gravitational
collapse of structure over time,
we measure both the nonlinear and stochastic components and verify that this decomposition of the
two point clustering statistics is reproduced within the simulation.
We find that the net effect of the nonlinearity manifests itself as an approximate rotation of $\langle \theta|\delta\rangle$ away
from the linear theory prediction $-f_{\rm LT}\delta$, and which increases as a function of scale.
The scatter about this mean value corresponds to stochasticity, or variance, of $\theta$ around $\langle \theta|\delta\rangle$ and which is nonzero on all scales.
The stochastic contribution to the velocity divergence auto power spectrum is approximately 10\% at $k<0.2h$Mpc$^{-1}$
at $z=0$. The stochastic component of the velocity
divergence auto power increases to approximately 25\% by
$k = 0.45h$Mpc$^{-1}$.

We examine two scales in detail, $k = 0.1h$Mpc$^{-1}$ and $k=0.2h$Mpc$^{-1}$, and find that
the scatter around the mean value $\langle \theta|\delta\rangle$ is nonzero and increases
with increasing wavenumber k. At both scales we find that
for $\delta > 0(<0)$ the mean relation $\langle \theta|\delta\rangle$  is larger (smaller)
then the linear theory prediction.
We find that both of these trends for the stochastic relation and nonlinearity are visible
at higher redshifts, $z=0.4$ and $z=1$ but with a reduced level of stochasticity overall due to less nonlinear growth at high
redshifts.
Using a halo sample with $M \le 5 \times 10^{12}M_\odot h^{-1}$
from the MultiDark simulation  we find  that both the stochasticity
and nonlinearity in the $\theta - \delta$ relation are larger for halos  compared to the dark
matter. We find that the stochastic component of the two point
function $\langle \theta \theta\rangle$
is significant and approximately a constant
fraction (15\%) at $k < 0.25h$Mpc$^{-1}$ at both $z = 0$ and
$z = 1$.

The relation with perturbative methods was also explored, and a computation of the velocity divergence auto $\langle \theta \theta\rangle$
and cross $\langle \theta \delta \rangle$ power spectra using one loop standard perturbation theory reveal that at this level the standard perturbation
theory prediction and the formalism in this paper are not equivalent. Even on large scales, $k < 0.05h$Mpc$^{-1}$, where the perturbation theory predictions match the measured power spectra from the simulations to  $\sim 5$\% there is no simple correspondence between the mode coupling terms $P_{13}$, which are negative, to the stochastic power spectra $\langle \alpha \delta \rangle$ and $\langle \alpha \alpha\rangle$.

Using an expression for $\theta$ computed from second order Lagrangian perturbation theory we find that the rotation of the conditional mean $\langle \theta |\delta \rangle$ away from the linear theory prediction is well described by the conditional expectation $\langle \theta |\delta \rangle_{\rm LPT}$ from 2LPT on scales  $k < 0.1 h$Mpc$^{-1}$. This indicates that the nonlinear components in the formalism can be described through the inclusion of tidal effects of the gravitational field at second order.

The central features discussed also have an impact on the extraction of the linear theory growth rate from models of two point functions in redshift space given the level of
 non-zero stochasticity which we have measured. It is common practice to try to extract a measurement of the linear growth rate using the $\mu^2$ and $\mu^4$ dependence of the measured galaxy power spectrum in redshift space. We highlight that, in the presence of either nonlinearity or a stochastic relation, the correspondence between the coefficients of $\mu^2$ and $\mu^4$, and $f_{\rm LT}$ is no longer so simple and a more involved treatment is required.

\section{Acknowledgements}

EJ would like to thank Roman Scoccimarro, Scott Dodelson and Andrey Kravtsov for helpful
discussions. 
EJ is supported 
by Fermi Research Alliance, LLC under 
the U.S. Department of Energy under contract No. DE-AC02-07CH11359.
 DJ is supported by the Royal Society.
We are grateful to Baojiu Li for allowing us to use the dark matter simulations presented in this study
which were performed on the ICC Cosmology Machine, which is part of
the DiRAC Facility jointly funded by STFC, the Large Facilities Capital Fund of BIS, and Durham University, and to
 the University of Chicago Research Computing Center for assistance with the calculations carried out in this work.
The MultiDark Database used in this paper and the web application providing online access to it were constructed as part of the activities of the German Astrophysical Virtual Observatory as result of a collaboration between the Leibniz-Institute for Astrophysics Potsdam (AIP) and the Spanish MultiDark Consolider Project CSD2009-00064. 
The MultiDark simulations were run on the NASA's Pleiades supercomputer at the NASA Ames Research Center.

\bibliographystyle{mn2e}
\bibliography{thebibliography}

\begin{thebibliography}{}

\bibitem[\protect\citeauthoryear{{Baumgart} \& {Fry}}{{Baumgart} \&
  {Fry}}{1991}]{1991ApJ...375...25B}
{Baumgart} D.~J.,  {Fry} J.~N.,  1991, \apj, 375, 25

\bibitem[\protect\citeauthoryear{{Bernardeau}, {Chodorowski}, {{\L}okas},
  {Stompor} \& {Kudlicki}}{{Bernardeau} et~al.}{1999}]{1999MNRAS.309..543B}
{Bernardeau} F.,  {Chodorowski} M.~J.,  {{\L}okas} E.~L.,  {Stompor} R.,
  {Kudlicki} A.,  1999, \mnras, 309, 543

\bibitem[\protect\citeauthoryear{{Bernardeau}, {Colombi}, {Gazta{\~n}aga} \&
  {Scoccimarro}}{{Bernardeau} et~al.}{2002}]{2002PhR...367....1B}
{Bernardeau} F.,  {Colombi} S.,  {Gazta{\~n}aga} E.,    {Scoccimarro} R.,
  2002, \physrep, 367, 1

\bibitem[\protect\citeauthoryear{{Bernardeau} \& {van de
  Weygaert}}{{Bernardeau} \& {van de Weygaert}}{1996}]{1996MNRAS.279..693B}
{Bernardeau} F.,  {van de Weygaert} R.,  1996, \mnras, 279, 693

\bibitem[\protect\citeauthoryear{{Beutler}, {Saito}, {Seo}, {Brinkmann},
  {Dawson}, {Eisenstein}, {Font-Ribera}, {Ho}, {McBride}, {Montesano},
  {Percival}, {Ross}, {Ross}, {Samushia}, {Schlegel}, {S{\'a}nchez}, {Tinker}
  \& {Weaver}}{{Beutler} et~al.}{2014}]{2014MNRAS.443.1065B}
{Beutler} F.,  {Saito} S.,  {Seo} H.-J.,  {Brinkmann} J.,  {Dawson} K.~S.,
  {Eisenstein} D.~J.,  {Font-Ribera} A.,  {Ho} S.,  {McBride} C.~K.,
  {Montesano} F.,  {Percival} W.~J.,  {Ross} A.~J.,  {Ross} N.~P.,  {Samushia}
  L.,  {Schlegel} D.~J.,  {S{\'a}nchez} A.~G.,  {Tinker} J.~L.,    {Weaver}
  B.~A.,  2014, \mnras, 443, 1065

\bibitem[\protect\citeauthoryear{{Biagetti}, {Desjacques}, {Kehagias} \&
  {Riotto}}{{Biagetti} et~al.}{2014}]{2014PhRvD..90j3529B}
{Biagetti} M.,  {Desjacques} V.,  {Kehagias} A.,    {Riotto} A.,  2014, \prd,
  90, 103529

\bibitem[\protect\citeauthoryear{{Bianchi}, {Chiesa} \& {Guzzo}}{{Bianchi}
  et~al.}{2015}]{2015MNRAS.446...75B}
{Bianchi} D.,  {Chiesa} M.,    {Guzzo} L.,  2015, \mnras, 446, 75

\bibitem[\protect\citeauthoryear{{Blake} et~al.,}{{Blake}
  et~al.}{2011}]{2011MNRAS.415.2876B}
{Blake} C.,  et~al., 2011, \mnras, 415, 2876

\bibitem[\protect\citeauthoryear{{Bonoli} \& {Pen}}{{Bonoli} \&
  {Pen}}{2009}]{2009MNRAS.396.1610B}
{Bonoli} S.,  {Pen} U.~L.,  2009, \mnras, 396, 1610

\bibitem[\protect\citeauthoryear{{Bouchet}}{{Bouchet}}{1996}]{1996dmu..conf..565B}
{Bouchet} F.~R.,  1996, in {Bonometto} S.,  {Primack} J.~R.,   {Provenzale} A.,
   eds, Dark Matter in the Universe {Introductory Overview of Eulerian and
  Lagrangian Perturbation Theories}.
p.~565

\bibitem[\protect\citeauthoryear{{Bouchet}, {Colombi}, {Hivon} \&
  {Juszkiewicz}}{{Bouchet} et~al.}{1995}]{1995A&A...296..575B}
{Bouchet} F.~R.,  {Colombi} S.,  {Hivon} E.,    {Juszkiewicz} R.,  1995, \aap,
  296, 575

\bibitem[\protect\citeauthoryear{{Cautun} \& {van de Weygaert}}{{Cautun} \&
  {van de Weygaert}}{2011}]{2011ascl.soft05003C}
{Cautun} M.~C.,  {van de Weygaert} R., , 2011, {The DTFE public software: The
  Delaunay Tessellation Field Estimator code}

\bibitem[\protect\citeauthoryear{{Chodorowski}}{{Chodorowski}}{1998}]{1998astro.ph..6087C}
{Chodorowski} M.,  1998, ArXiv Astrophysics e-prints

\bibitem[\protect\citeauthoryear{Cimatti et~al.,}{Cimatti
  et~al.}{2009}]{Cimatti:2009is}
Cimatti A.,  et~al., 2009, arXiv/0912.0914

\bibitem[\protect\citeauthoryear{{Colombi}, {Chodorowski} \&
  {Teyssier}}{{Colombi} et~al.}{2007}]{2007MNRAS.375..348C}
{Colombi} S.,  {Chodorowski} M.~J.,    {Teyssier} R.,  2007, \mnras, 375, 348

\bibitem[\protect\citeauthoryear{{Crocce} \& {Scoccimarro}}{{Crocce} \&
  {Scoccimarro}}{2006}]{2006PhRvD..73f3519C}
{Crocce} M.,  {Scoccimarro} R.,  2006, \prd, 73, 063519

\bibitem[\protect\citeauthoryear{{Crocce} \& {Scoccimarro}}{{Crocce} \&
  {Scoccimarro}}{2008}]{2008PhRvD..77b3533C}
{Crocce} M.,  {Scoccimarro} R.,  2008, \prd, 77, 023533

\bibitem[\protect\citeauthoryear{{Crocce}, {Scoccimarro} \&
  {Bernardeau}}{{Crocce} et~al.}{2012}]{2012MNRAS.427.2537C}
{Crocce} M.,  {Scoccimarro} R.,    {Bernardeau} F.,  2012, \mnras, 427, 2537

\bibitem[\protect\citeauthoryear{{Dekel} \& {Lahav}}{{Dekel} \&
  {Lahav}}{1999}]{1999ApJ...520...24D}
{Dekel} A.,  {Lahav} O.,  1999, \apj, 520, 24

\bibitem[\protect\citeauthoryear{{Eisenstein} \& {DESI
  Collaboration}}{{Eisenstein} \& {DESI
  Collaboration}}{2015}]{2015AAS...22533605E}
{Eisenstein} D.,  {DESI Collaboration} 2015, in American Astronomical Society
  Meeting Abstracts Vol.~225 of American Astronomical Society Meeting
  Abstracts, {The Dark Energy Spectroscopic Instrument (DESI): Science from the
  DESI Survey}.
p. 336.05

\bibitem[\protect\citeauthoryear{{Feldman}, {Kaiser} \& {Peacock}}{{Feldman}
  et~al.}{1994}]{1994ApJ...426...23F}
{Feldman} H.~A.,  {Kaiser} N.,    {Peacock} J.~A.,  1994, \apj, 426, 23

\bibitem[\protect\citeauthoryear{{Gramann}}{{Gramann}}{1993}]{1993ApJ...405L..47G}
{Gramann} M.,  1993, \apjl, 405, L47

\bibitem[\protect\citeauthoryear{{Guzzo} et~al.,}{{Guzzo}
  et~al.}{2008}]{2008Natur.451..541G}
{Guzzo} L.,  et~al., 2008, \nat, 451, 541

\bibitem[\protect\citeauthoryear{{Hamilton}}{{Hamilton}}{1998}]{1998ASSL..231..185H}
{Hamilton} A.~J.~S.,  1998, in {Hamilton} D.,  ed., The Evolving Universe
  Vol.~231 of Astrophysics and Space Science Library, {Linear Redshift
  Distortions: a Review}.
p.~185

\bibitem[\protect\citeauthoryear{{Hockney} \& {Eastwood}}{{Hockney} \&
  {Eastwood}}{1988}]{1988csup.book.....H}
{Hockney} R.~W.,  {Eastwood} J.~W.,  1988, {Computer simulation using
  particles}

\bibitem[\protect\citeauthoryear{{Jennings}}{{Jennings}}{2012}]{2012MNRAS.427L..25J}
{Jennings} E.,  2012, \mnras, 427, L25

\bibitem[\protect\citeauthoryear{{Jennings}, {Baugh} \& {Hatt}}{{Jennings}
  et~al.}{2015}]{2015MNRAS.446..793J}
{Jennings} E.,  {Baugh} C.~M.,    {Hatt} D.,  2015, \mnras, 446, 793

\bibitem[\protect\citeauthoryear{{Jennings}, {Baugh}, {Li}, {Zhao} \&
  {Koyama}}{{Jennings} et~al.}{2012}]{2012MNRAS.425.2128J}
{Jennings} E.,  {Baugh} C.~M.,  {Li} B.,  {Zhao} G.-B.,    {Koyama} K.,  2012,
  \mnras, 425, 2128

\bibitem[\protect\citeauthoryear{{Jennings}, {Baugh} \& {Pascoli}}{{Jennings}
  et~al.}{2011}]{2011MNRAS.410.2081J}
{Jennings} E.,  {Baugh} C.~M.,    {Pascoli} S.,  2011, MNRAS, 410, 2081, 410,
  2081

\bibitem[\protect\citeauthoryear{{Kaiser}}{{Kaiser}}{1987}]{1987MNRAS.227....1K}
{Kaiser} N.,  1987, \mnras, 227, 1

\bibitem[\protect\citeauthoryear{{Kitaura}, {Angulo}, {Hoffman} \&
  {Gottl{\"o}ber}}{{Kitaura} et~al.}{2012}]{2012MNRAS.425.2422K}
{Kitaura} F.-S.,  {Angulo} R.~E.,  {Hoffman} Y.,    {Gottl{\"o}ber} S.,  2012,
  \mnras, 425, 2422

\bibitem[\protect\citeauthoryear{{Klypin} \& {Holtzman}}{{Klypin} \&
  {Holtzman}}{1997}]{1997astro.ph.12217K}
{Klypin} A.,  {Holtzman} J.,  1997, ArXiv Astrophysics e-prints

\bibitem[\protect\citeauthoryear{{Kravtsov} \& {Klypin}}{{Kravtsov} \&
  {Klypin}}{1999}]{1999ApJ...520..437K}
{Kravtsov} A.~V.,  {Klypin} A.~A.,  1999, \apj, 520, 437

\bibitem[\protect\citeauthoryear{{Kwan}, {Lewis} \& {Linder}}{{Kwan}
  et~al.}{2012}]{2012ApJ...748...78K}
{Kwan} J.,  {Lewis} G.~F.,    {Linder} E.~V.,  2012, \apj, 748, 78

\bibitem[\protect\citeauthoryear{{Li}, {Hellwing}, {Koyama}, {Zhao}, {Jennings}
  \& {Baugh}}{{Li} et~al.}{2013}]{2013MNRAS.428..743L}
{Li} B.,  {Hellwing} W.~A.,  {Koyama} K.,  {Zhao} G.-B.,  {Jennings} E.,
  {Baugh} C.~M.,  2013, \mnras, 428, 743

\bibitem[\protect\citeauthoryear{{Li}, {Zhao}, {Teyssier} \& {Koyama}}{{Li}
  et~al.}{2012}]{2012JCAP...01..051L}
{Li} B.,  {Zhao} G.-B.,  {Teyssier} R.,    {Koyama} K.,  2012, JCAP, 1, 51

\bibitem[\protect\citeauthoryear{{Melott}, {Buchert} \& {Weib}}{{Melott}
  et~al.}{1995}]{1995A&A...294..345M}
{Melott} A.~L.,  {Buchert} T.,    {Weib} A.~G.,  1995, \aap, 294, 345

\bibitem[\protect\citeauthoryear{{Neyrinck}, {Szapudi} \& {Szalay}}{{Neyrinck}
  et~al.}{2009}]{2009ApJ...698L..90N}
{Neyrinck} M.~C.,  {Szapudi} I.,    {Szalay} A.~S.,  2009, \apjl, 698, L90

\bibitem[\protect\citeauthoryear{{Peacock} \& {Dodds}}{{Peacock} \&
  {Dodds}}{1994}]{1994MNRAS.267.1020P}
{Peacock} J.~A.,  {Dodds} S.~J.,  1994, \mnras, 267, 1020

\bibitem[\protect\citeauthoryear{{Peacock} et~al.,}{{Peacock}
  et~al.}{2001}]{2001Natur.410..169P}
{Peacock} J.~A.,  et~al., 2001, \nat, 410, 169

\bibitem[\protect\citeauthoryear{{Percival} \& {White}}{{Percival} \&
  {White}}{2009}]{2009MNRAS.393..297P}
{Percival} W.~J.,  {White} M.,  2009, \mnras, 393, 297

\bibitem[\protect\citeauthoryear{{Prada}, {Klypin}, {Cuesta}, {Betancort-Rijo}
  \& {Primack}}{{Prada} et~al.}{2012}]{2012MNRAS.423.3018P}
{Prada} F.,  {Klypin} A.~A.,  {Cuesta} A.~J.,  {Betancort-Rijo} J.~E.,
  {Primack} J.,  2012, \mnras, 423, 3018

\bibitem[\protect\citeauthoryear{{Pueblas} \& {Scoccimarro}}{{Pueblas} \&
  {Scoccimarro}}{2009}]{2009PhRvD..80d3504P}
{Pueblas} S.,  {Scoccimarro} R.,  2009, \prd, 80, 043504

\bibitem[\protect\citeauthoryear{{Reid} et~al.,}{{Reid}
  et~al.}{2012}]{2012MNRAS.426.2719R}
{Reid} B.~A.,  et~al., 2012, \mnras, 426, 2719

\bibitem[\protect\citeauthoryear{{Reid} \& {White}}{{Reid} \&
  {White}}{2011}]{2011MNRAS.417.1913R}
{Reid} B.~A.,  {White} M.,  2011, \mnras, 417, 1913

\bibitem[\protect\citeauthoryear{{Riebe}, {Partl}, {Enke}, {Forero-Romero},
  {Gottloeber}, {Klypin}, {Lemson}, {Prada}, {Primack}, {Steinmetz} \&
  {Turchaninov}}{{Riebe} et~al.}{2011}]{2011arXiv1109.0003R}
{Riebe} K.,  {Partl} A.~M.,  {Enke} H.,  {Forero-Romero} J.,  {Gottloeber} S.,
  {Klypin} A.,  {Lemson} G.,  {Prada} F.,  {Primack} J.~R.,  {Steinmetz} M.,
  {Turchaninov} V.,  2011, ArXiv e-prints

\bibitem[\protect\citeauthoryear{{S{\'a}nchez}, {Crocce}, {Cabr{\'e}}, {Baugh}
  \& {Gazta{\~n}aga}}{{S{\'a}nchez} et~al.}{2009}]{Sanchez:2009jq}
{S{\'a}nchez} A.~G.,  {Crocce} M.,  {Cabr{\'e}} A.,  {Baugh} C.~M.,
  {Gazta{\~n}aga} E.,  2009, \mnras, 400, 1643

\bibitem[\protect\citeauthoryear{{Sato} \& {Matsubara}}{{Sato} \&
  {Matsubara}}{2013}]{2013PhRvD..87l3523S}
{Sato} M.,  {Matsubara} T.,  2013, \prd, 87, 123523

\bibitem[\protect\citeauthoryear{{Schaap}}{{Schaap}}{2007}]{2007PhDT.......433S}
{Schaap} W.~E.,  2007, PhD thesis, Kapteyn Astronomical Institute

\bibitem[\protect\citeauthoryear{Scoccimarro}{Scoccimarro}{2004}]{Scoccimarro:2004tg}
Scoccimarro R.,  2004, \prd, 70, 083007

\bibitem[\protect\citeauthoryear{{Seljak} \& {McDonald}}{{Seljak} \&
  {McDonald}}{2011}]{2011JCAP...11..039S}
{Seljak} U.,  {McDonald} P.,  2011, JCAP, 11, 39

\bibitem[\protect\citeauthoryear{{Seljak} \& {Warren}}{{Seljak} \&
  {Warren}}{2004}]{2004MNRAS.355..129S}
{Seljak} U.,  {Warren} M.~S.,  2004, \mnras, 355, 129

\bibitem[\protect\citeauthoryear{{Spergel} et~al.,}{{Spergel}
  et~al.}{2013}]{2013arXiv1305.5422S}
{Spergel} D.,  et~al., 2013, ArXiv e-prints

\bibitem[\protect\citeauthoryear{{Taruya}, {Nishimichi} \&
  {Bernardeau}}{{Taruya} et~al.}{2013}]{2013PhRvD..87h3509T}
{Taruya} A.,  {Nishimichi} T.,    {Bernardeau} F.,  2013, \prd, 87, 083509

\bibitem[\protect\citeauthoryear{{Teyssier}}{{Teyssier}}{2002}]{2002A&A...385..337T}
{Teyssier} R.,  2002, \aap, 385, 337

\bibitem[\protect\citeauthoryear{{Zhang}, {Zheng} \& {Jing}}{{Zhang}
  et~al.}{2014}]{2014arXiv1405.7125Z}
{Zhang} P.,  {Zheng} Y.,    {Jing} Y.,  2014, ArXiv e-prints

\bibitem[\protect\citeauthoryear{{Zheng}, {Zhang} \& {Jing}}{{Zheng}
  et~al.}{2014}]{2014arXiv1410.1256Z}
{Zheng} Y.,  {Zhang} P.,    {Jing} Y.,  2014, ArXiv e-prints

\end{thebibliography}

\end{document}